\documentclass[final,5p,times,twocolumn]{elsarticle}

\renewcommand{\normalsize}{\fontsize{8pt}{9.6pt}\selectfont}
\renewcommand{\small}{\fontsize{7pt}{8.4pt}\selectfont}
\renewcommand{\footnotesize}{\fontsize{6pt}{7.2pt}\selectfont}
\renewcommand{\scriptsize}{\fontsize{5pt}{6pt}\selectfont}

\usepackage{amssymb}
\usepackage{amsmath}
\usepackage{url}

\usepackage{amssymb}
\usepackage{amsmath}
\usepackage{url}

\journal{Blockchain: Research and Applications}

\begin{document}

\begin{frontmatter}



\title{Systematizing DAO Governance: Design, Effective Decentralization,
Resilience, and Economic Outcomes}

\author[1]{Yash Madhwal\corref{cor1}}
\ead{yash.madhwal@skoltech.ru}
\author[2]{Vladimir Lebedev}
\author[3]{Alexey Karanyuk}
\author[1]{Yury Yanovich}

\cortext[cor1]{Corresponding author}

\affiliation[1]{
    organization={Skolkovo Institute of Science and Technology},
    city={Moscow},
    country={Russia}
}

\affiliation[2]{
    organization={National Research University Higher School of Economics},
    city={Moscow},
    country={Russia}
}

\affiliation[3]{
    organization={Giants Tokenomics Lab},
    city={Florianópolis},
    country={Brazil}
}

\begin{abstract}
Decentralized Autonomous Organizations (DAOs) combine formal governance rules with delegation, voting thresholds, execution constraints, emergency authority, and external dependencies. This paper systematizes DAO governance as a configuration of authority and complements the Systematization of Knowledge with an empirical analysis of 54 Tally-indexed DAOs. We find that formal openness often coexists with concentrated effective authority: among the 52 DAOs with delegate-level data, the largest delegate controls a median of about 42\% of delegated voting power, and the ten largest delegates control about 98\%. Quorum analysis shows that, among DAOs with observable quorum thresholds, about two-thirds can satisfy quorum through their largest delegate alone, although this does not imply coordination or capture. Token-market analyses indicate that market scale is more consistently associated with governance breadth than with voting-power concentration, while static governance measures add little explanatory power for token returns once Bitcoin exposure is controlled for; proposal-event estimates are negative but weaken under dependence corrections. The results argue for evaluating DAO decentralization through realized authority distribution, exercise, and constraint, rather than through binary labels or single composite scores.
\end{abstract}

\begin{highlights}
\item DAO governance is framed as a configuration of authority
\item Largest delegate controls median 42\% of voting power in 52 DAOs
\item Ten largest delegates control median 98\% of delegated voting power
\item Two-thirds of DAOs with observable quorum reach quorum via largest delegate
\item Static governance metrics show weak return associations after Bitcoin exposure
\end{highlights}


\end{frontmatter}




\section{Introduction}
\label{sec:introduction}

Decentralized Autonomous Organizations (DAOs) have evolved into blockchain-based organizational systems in which collective decision-making is combined with programmable execution and community-based governance \cite{hsieh2018bitcoin,hassan2021dao,santana2022dao,elfaqir2020overview,rikken2023insouts}. This organizational form has been explored not only for broad community treasuries but also for special-purpose financing, where governance rules determine how funds are proposed, approved, and disbursed \cite{Bischof2022}. The distributed-ledger and smart-contract foundations of these systems originate in the architectures introduced by Bitcoin and Ethereum \cite{Nakamoto2008,Buterin2014}. Unlike conventional organizations, in which formal decision authority is typically exercised through managerial or legal hierarchies, DAOs coordinate participants through a combination of blockchain-based rules, smart contracts, and collective governance processes \cite{appel2023control}. Governance is commonly exercised through token-weighted voting, while many DAO systems also allow token holders to delegate voting rights to representatives who vote on their behalf \cite{sun2024makerdao,weidener2025delegated}. Consequently, DAO governance extends beyond voting itself and encompasses agenda setting, representation, decision thresholds, execution constraints, emergency authority, and other on-chain and off-chain mechanisms that shape collective decision-making \cite{reijers2021code}.

DAO governance is therefore better understood as a \emph{socio-technical system} than as a purely on-chain voting mechanism \cite{santana2022dao,reijers2021code,beck2018blockchain,vanpelt2021framework}. Proposal and voting contracts, quorum requirements, timelocks, and automated execution interact with off-chain deliberation, signaling processes, delegated representation, governance forums, and specialized operational or security actors \cite{zhu2024governance,liu2022principles,schadler2023decisionmaking}. These components jointly determine who can place issues on the agenda, whose preferences influence outcomes, how approved decisions progress toward execution, and under what conditions exceptional intervention is possible. Governance design consequently involves recurring trade-offs between openness and gatekeeping, representation and concentration, decision accessibility and coalition breadth, and decentralized legitimacy and emergency responsiveness \cite{feichtinger2024sokdaos,honkanen2021buildingblocks}.

\subsection*{From formal rules to effective authority}

A central challenge is that \emph{formal decentralization} and \emph{effective decentralization} need not coincide. Decentralization has technical, organizational, political, and economic dimensions that should not be collapsed into a single architectural property \cite{bodo2021decentralisation,defilippi2016politics,defilippi2020confidence}. A DAO may formally permit a broad population of token holders or delegates to participate while effective decision-making authority remains concentrated among a comparatively small number of actors. Empirical studies have documented substantial concentration of voting power despite formally decentralized participation mechanisms \cite{goldberg2023metaverse,fritsch2024votingpower,han2026dao,sun2024makerdao,feichtinger2024shortcomings}. Delegation can reduce the participation burden on individual token holders and support specialized representation, but it can also redistribute effective voting authority toward a smaller set of active or highly delegated representatives \cite{gilson2024democratic,weidener2025delegated}. Similarly, a formal quorum requirement may appear demanding in absolute terms while remaining arithmetically satisfiable by one or a few highly weighted delegates. Open participation and transparent governance records should therefore not, by themselves, be interpreted as evidence that effective decision-making authority is broadly distributed.

This distinction is especially important because DAO governance is unusually observable. Proposal histories, voting records, delegated voting power, governance parameters, and execution activity can often be reconstructed from public blockchain records or standardized governance interfaces such as Snapshot and Tally \cite{snapshotdocs,tallydocs,Tally2024}. This observability enables researchers to connect formal governance architecture with quantitative measures of realized authority, including participation, voting-power concentration, and proposal-level decision behavior \cite{penacalvin2024concentration,han2026dao}. At the same time, transparency should not be conflated with decentralization: public records make authority easier to observe, but they do not determine how broadly that authority is distributed.

Existing research approaches these questions through several partially distinct strands. One strand examines the institutional and technical design of on-chain and off-chain governance \cite{reijers2021code,zhu2024governance,beck2018blockchain,vanpelt2021framework,liu2022principles,azouvi2019governance}. A second studies voting-power concentration, delegation, participation, voter behavior, and the distribution of effective control \cite{goldberg2023metaverse,gilson2024democratic,fritsch2024votingpower,weidener2025delegated,arroyo2022daoAnalyzer,rikken2023impacts}. Security-oriented research considers governance capture, procedural manipulation, privileged roles, and governance-related attack surfaces \cite{feichtinger2024sokdaos,feichtinger2024shortcomings}. Research on commons governance and open-source production emphasizes that formal rules coexist with community practices and contribution networks \cite{rozas2021ostrom,vanvulpen2024oss}. A further strand examines whether governance characteristics are associated with organizational development, valuation, or token-market outcomes \cite{han2026dao,bellavitis2025voting,lomonaco2025distributed}. These literatures provide substantial evidence on individual dimensions of DAO governance, but comparatively few studies jointly examine governance design, the realized distribution of decision authority, economic outcomes, and governance behavior under stress within a common analytical framework.

The contribution is therefore an integrative account of DAO governance as a configuration of authority. It links how formal rules allocate decision rights, how those rights are distributed and exercised in deployed systems, and how the resulting authority arrangement relates to token-market outcomes and response capacity under stress.

\subsection*{Research questions and contributions}

To address this fragmentation, this paper combines a Systematization of Knowledge (SoK) with a cross-DAO empirical analysis. The paper organizes governance architecture, delegation, participation, security-relevant constraints, and economic performance as interconnected layers of authority. First, it systematizes DAO governance around six recurring authority layers: agenda access, voting and delegation, decision thresholds, execution constraints, emergency authority, and external dependencies. These layers are not intended as an exhaustive inventory of all governance mechanisms; rather, they identify the stable dimensions along which authority is allocated, concentrated, exercised, and constrained across deployed DAOs. Second, it measures whether that formal architecture translates into distributed effective authority through observable indicators of voting-power concentration, participation, and quorum-sufficiency structure. Third, it examines whether these governance characteristics are associated with token-market scale and token returns. Finally, it analyzes how concentration, procedural safeguards, privileged authority, and external dependencies shape governance exposure and response under adversarial or emergency conditions.

The empirical component combines governance information obtained through Tally with token-market information from CryptoCompare. The sample is constructed from the full set of organizations indexed by Tally rather than from a manually selected set of prominent DAOs. After applying explicit eligibility, governance-token, mainnet, duplication, and token-matching rules, the final governance universe contains 54 DAOs identified by unique mainnet governance-token contracts. Delegate-level data required for voting-power concentration are available for 52 DAOs, daily USD price histories for 46, and at least 12 months of return history for 45. Because individual analyses require different combinations of governance, quorum, and market information, estimation samples vary across specifications rather than being restricted to a single common subset.

Rather than reducing decentralization to a single indicator, we distinguish three complementary dimensions of effective governance. Voting-power concentration measures how delegated authority is distributed among representatives; participation captures the extent to which governance rights are exercised; and the Effective Quorum Coalition measures how many of the largest delegates are arithmetically sufficient to satisfy a formal quorum requirement. These dimensions describe different properties of authority and are therefore analyzed separately rather than combined into a universal decentralization score.

The economic analysis similarly separates token-market scale from token returns. This distinction is important because governance may be associated with organizational breadth, adoption, or valuation without generating predictable short-run asset returns. The corporate-governance literature provides a useful conceptual precedent: agency problems, ownership concentration, and complementary governance mechanisms shape control and valuation, while the association between governance indices and abnormal returns may weaken as information becomes incorporated into prices \cite{Jensen1976,Shleifer1997,Gompers2003,Cremers2005,Bebchuk2009,Bebchuk2013}. DAO governance provides a particularly observable setting in which to revisit this distinction. Accordingly, we examine whether governance characteristics are associated with token-market scale and whether they provide additional explanatory power for token returns after accounting for broader crypto-market movements. These analyses are explicitly observational and do not identify a causal pathway from governance to valuation or returns.

This integrated perspective leads to four research questions:

\begin{description}

\item[\textbf{RQ1 --- Governance Design.}] How do deployed DAOs structure agenda access, voting and delegation, decision thresholds, execution constraints, emergency authority, and external dependencies?

\item[\textbf{RQ2 --- Effective Decentralization.}] To what extent does formal DAO governance translate into effective decentralization when measured through voting-power concentration, participation, and quorum coalition structure?

\item[\textbf{RQ3 --- Economic Outcomes.}] Are observable governance characteristics associated with DAO token-market scale or token returns after accounting for systematic crypto-market exposure?

\item[\textbf{RQ4 --- Governance Under Stress.}] How do concentration, privileged authority, procedural safeguards, and external dependencies shape governance exposure and resilience under adversarial or emergency conditions?

\end{description}

Here, resilience refers to the design capacity of governance arrangements to absorb, escalate, pause, replace, or reverse decisions under adversarial or emergency conditions. It is analyzed through mechanism design and illustrative incidents, rather than through population-level estimates of failure frequency or response latency.

To answer these questions, the paper makes three principal contributions.

\begin{itemize}

\item \textbf{A unified governance design space.} We systematize DAO governance around recurring design primitives spanning agenda access, voting and delegation, decision thresholds, execution constraints, emergency authority, and external dependencies. The framework connects routine collective decision-making with privileged, safety-critical, and dependency-sensitive control mechanisms and maps governance design to recurring trade-offs and failure exposures.

\item \textbf{An empirical assessment of effective decentralization.} We complement formal governance architecture with quantitative measures of voting-power concentration, participation, and effective quorum coalition structure. This allows us to distinguish nominal participation capacity from the realized distribution of decision-making authority in deployed governance systems.

\item \textbf{An integrated assessment of governance and observable outcomes.} We relate effective governance characteristics to token-market scale and token returns and connect these empirical observations to the failure exposures and safeguards identified by the systematization. This separates the institutional and operational significance of governance from the stronger claim that governance generates predictable short-run token-market performance.

\end{itemize}

\subsection*{Evidence map}

The relation among the six authority layers, the empirical measures, and the interpretation boundaries is summarized in Table~\ref{tab:rq-roadmap}.

\begin{table*}[htbp]
\centering
\small
\caption{Roadmap from research questions to evidence and interpretation.}
\label{tab:rq-roadmap}
\begin{tabular}{@{}p{0.12\textwidth}p{0.20\textwidth}p{0.24\textwidth}p{0.32\textwidth}@{}}
\hline
\textbf{Question} & \textbf{Evidence base} & \textbf{Key measurement / object} & \textbf{Interpretation boundary} \\
\hline
RQ1: Governance design &
SoK corpus and protocol documentation &
Six recurring authority layers &
Identifies recurring mechanisms and trade-offs; not an exhaustive census of all DAO designs \\
\hline
RQ2: Effective decentralization &
54 Tally-indexed DAOs; 52 with delegate-level data &
HHI, top-$k$ shares, participation, EQC (arithmetic quorum sufficiency) &
Measures realized voting-power structure; snapshot-based and not causal evidence of capture \\
\hline
RQ3: Economic outcomes &
Token-market subsamples &
Live-token FDV, BTC-relative returns, proposal-event estimates &
Observational associations; distinguishes market scale from governance quality and short-run pricing \\
\hline
RQ4: Governance under stress &
Illustrative incidents and design analysis &
Emergency authority, timelocks, oracle and infrastructure dependencies &
Maps structural exposure and response capacity; does not support population-level incident-frequency claims \\
\hline
\end{tabular}
\end{table*}

A preview of the main empirical findings is as follows. Effective voting authority is substantially concentrated: among the 52 DAOs with delegate-level data, the median largest delegate controls about 42\% of delegated voting power, and the ten largest delegates control about 98\%. Among the 38 DAOs whose observed delegation distribution reaches a valid quorum, the median Effective Quorum Coalition is one, and about two-thirds exhibit single-delegate quorum sufficiency. These findings indicate structural concentration, subject to the arithmetic-coalition interpretation of EQC.

The economic evidence is more qualified. Token-market scale is more consistently associated with the breadth of the delegate and token-holder bases than with independently lower voting-power concentration. Static governance characteristics do not robustly explain average token returns, while Bitcoin exposure accounts for substantially more monthly token-return variation. Proposal-completion events show negative returns relative to Bitcoin, but the evidence weakens under dependence corrections.

Taken together, these findings motivate a layered interpretation of DAO decentralization. Formal governance determines where authority \emph{can} reside; effective-decentralization measures reveal where voting authority \emph{does} reside; and governance under stress reveals how exceptional authority can be exercised when ordinary collective procedures are insufficient. The central analytical question is therefore not simply whether a DAO is decentralized, but where authority is located, how concentrated and actively exercised it is, how broadly it must be combined to make decisions, and which safeguards constrain its use across the governance lifecycle.


\section{Methodology}
\label{sec:methodology}

This study combines a SoK with a cross-DAO empirical analysis. The two components address complementary aspects of DAO governance. The SoK component examines how governance authority is formally structured across six recurring layers: agenda access, voting and delegation, decision thresholds, execution constraints, emergency authority, and external dependencies. The empirical component examines how this formal architecture translates into observable distributions of effective decision-making authority and whether these governance characteristics are associated with token-market scale and token returns.

We therefore distinguish two evidence layers throughout the study. The first is a \emph{governance-design evidence layer}, constructed from academic literature, protocol documentation, governance proposals and forums, on-chain execution records, and documented incidents. The second is an \emph{empirical governance layer}, constructed from standardized governance and market data and used to measure voting-power concentration, participation, quorum coalition structure, and economic outcomes.

This separation is important because the two evidence layers serve different inferential purposes. Protocol-level cases provide detailed evidence on mechanisms, institutional arrangements, and behavior under stress, whereas the quantitative dataset provides broader cross-sectional evidence on measurable governance structure. We therefore do not treat the protocol case corpus and the empirical DAO sample as a single common sample.

\subsection{SoK Scope, Evidence Strategy, and Case Selection}
\label{sec:sok-scope}

The systematization focuses on governance mechanisms that determine how authority is allocated, exercised, constrained, and monitored within DAOs. The scope includes both routine governance and exceptional control mechanisms. Routine governance encompasses agenda access, proposal formation, voting and delegation, decision thresholds, and execution constraints. Exceptional governance encompasses privileged safety roles, emergency intervention, governance capture, procedural manipulation, and dependencies whose failure may require governance intervention.

The evidence base combines academic and primary governance sources. Academic literature is used to establish existing conceptual frameworks, empirical findings, security models, and previously identified governance patterns. Primary governance evidence is used to determine how mechanisms are implemented and exercised in deployed protocols. These sources include public governance documentation, proposal archives, voting and delegation records, governance forums, execution traces, incident reports, and post-mortems. Governance interfaces and analytics platforms such as Snapshot, Tally, DeepDAO, and Karma provide additional access to standardized governance artifacts \cite{snapshotdocs,tallydocs,Tally2024,deepdao2025,karma}.

The qualitative comparison uses a purposive maximum-variation set of nine
mature governance systems: Uniswap, Aave, MakerDAO, Compound, Arbitrum,
Optimism, ENS, Gitcoin, and Nouns. A protocol was eligible when it satisfied
four criteria: (i) a sustained and publicly traceable governance process;
(ii) accessible proposal, voting, and execution artifacts; (iii) a mechanism
that adds variation to the comparison, such as delegation, staged execution,
or a specialized emergency actor; and (iv) sufficient documentation to avoid
inferring institutional properties from interfaces alone. The set spans
decentralized finance, infrastructure, public goods, identity, and
community-oriented governance. It is intended to expose recurring design
patterns, not to estimate their frequency in the DAO population. The broader
Tally sample used for the quantitative analysis is constructed independently
and is not restricted to these nine cases.

\subsection{Evidence Identification and Source Hierarchy}
\label{sec:literature-search}

The academic evidence was assembled as a structured, concept-centric SoK
corpus rather than as an exhaustive bibliometric review. This approach follows
established guidance that a review should organize evidence around concepts,
make its inclusion logic explicit, and synthesize relationships rather than
list studies sequentially \cite{webster2002review,pare2015reviews,okoli2015review}.
Retrieval began from foundational blockchain-governance frameworks addressing
decision rights, accountability, incentives, governance layers, and design
principles \cite{beck2018blockchain,vanpelt2021framework,
liu2022principles,honkanen2021buildingblocks}. It then incorporated DAO-focused
conceptual and empirical research on organizational form, delegation,
concentration, participation, value creation, and governance attacks
\cite{hassan2021dao,reijers2021code,zhu2024governance,
feichtinger2024sokdaos,weidener2025delegated,elfaqir2020overview,
rikken2023impacts,arroyo2022daoAnalyzer}. The corpus was expanded through
backward and forward citation chaining and targeted searches combining
``DAO'' or ``decentralized autonomous organization'' with ``governance'',
``voting'', ``delegation'', ``participation'', ``quorum'', ``voting power'',
``capture'', ``attack'', ``emergency'', ``oracle'', ``valuation'', and
``returns''. Evidence available through 5 October 2026 was considered.

A source entered the academic corpus when it contributed a governance model,
an empirical result, a security or failure analysis, or evidence connecting
governance to organizational or market outcomes. Sources that merely mentioned
DAOs, duplicated a stronger version of the same work, or lacked enough detail
to support a coded claim were excluded from the synthesis. Citation chaining
was stopped when additional sources reproduced an already represented
mechanism without changing the taxonomy or its boundary conditions. This retrieval strategy is concept-centric and designed to saturate recurring governance mechanisms, boundary conditions, and trade-offs. Inclusion logic, corpus size, and stopping rule are reported explicitly to support auditability. The corpus is therefore a structured SoK evidence base for framework synthesis, rather than a bibliometric population estimate. A concise search-protocol summary is provided in Appendix~\ref{app:search-protocol}.

The final core academic coding corpus contains 34 peer-reviewed journal or
conference studies. Protocol documentation, market-data documentation, and
general statistical or review-method sources are cited where relevant but are
not counted in this total. Appendix~\ref{app:sok-evidence-matrix} lists every
core study and records its governance dimension, method, scope, evidence, and
principal finding.

Protocol-level claims follow a fixed evidence hierarchy. Deployed contract
state and execution records take priority, followed by official governance
proposals, documentation, and post-mortems; peer-reviewed research is then
used for interpretation and comparison; third-party analytics and practitioner
commentary are used only as supporting evidence. When sources describe
different protocol versions, the claim is scoped to the identifiable version
or observation date rather than generalized to the protocol indefinitely.

\subsection{Governance Coding Framework}
\label{sec:coding}

To enable comparison across heterogeneous governance systems, evidence is coded into a common set of governance dimensions. The coding framework focuses on mechanisms that determine agenda access, preference aggregation, effective authority, execution, and exceptional control.

\begin{table}[t]
\centering
\caption{Governance dimensions used in the comparative coding framework.}
\label{tab:coding-framework}
\begin{tabular}{p{0.25\columnwidth} p{0.65\columnwidth}}
\hline
\textbf{Dimension} & \textbf{Illustrative coded attributes} \\
\hline

Agenda access &
Permissionless submission, voting-power threshold, sponsorship,
facilitator or forum gating \\

Voting &
Token-weighted, delegated, off-chain signaling, hybrid mechanisms \\

Delegation &
Direct voting, transferable delegation, representative delegates,
delegate concentration \\

Decision rules and thresholds &
Quorum, majority, supermajority, vote-differential requirements \\

Execution &
Immediate execution, timelock, staged execution,
security delay \\

Emergency authority &
Guardian, multisignature committee, security council,
other privileged actors \\

Participation &
Voter activity, delegate activity, participation incentives,
accountability requirements \\

Capture exposure &
Voting-power concentration, narrow quorum coalition,
agenda concentration \\

Safety controls &
Pause, freeze, circuit breaker, shutdown,
revocation or emergency override \\
\hline
\end{tabular}
\end{table}

The coding unit depends on the governance property under examination. Protocol-level architecture is coded at the DAO or governor level; proposal-process properties are coded at the governance-process level; and emergency behavior is coded at the incident level. This distinction avoids treating static institutional characteristics and individual governance events as equivalent observations.

Classifications are grounded in verifiable artifacts, including governance
contracts, protocol documentation, proposal text, voting outcomes, execution
transactions, and linked governance discussions. Secondary analytics and
practitioner commentary complement rather than substitute for primary
evidence. Conflicts are resolved using the hierarchy described in
Section~\ref{sec:literature-search}; if a source cannot be associated with a
governance version or an observation date, the corresponding classification is
qualified or omitted. The qualitative coding is used to derive mechanisms and
boundary conditions, not numerical prevalence estimates. We therefore do not
report an inter-rater agreement statistic for the case corpus.

\subsection{Comparative Governance Synthesis}
\label{sec:comparative-synthesis}

The coded evidence is synthesized comparatively rather than by treating any single DAO as representative of decentralized governance as a whole. The analysis searches for recurring governance primitives and trade-offs across protocols while preserving protocol-specific differences.

The comparison is organized around three questions. First, which governance mechanisms recur across otherwise heterogeneous DAOs? Second, how do different implementations redistribute effective authority among token holders, delegates, councils, and emergency actors? Third, which design choices repeatedly generate tensions between openness, legitimacy, responsiveness, and capture resistance?

Documented incidents are incorporated as targeted stress cases. Their purpose
is to examine how governance mechanisms behave under operational pressure, not
to estimate the population frequency of governance failures. The MakerDAO,
Compound, and Aave cases are therefore treated explicitly as illustrative
mechanism tests and are excluded from any frequency or effectiveness claim.

\subsection{Empirical Governance Dataset}
\label{sec:empirical-data}

The empirical component combines governance data collected from the Tally
(Cactus) GraphQL API with token-market data from CryptoCompare. Tally provides
organization, governor, governance-token, delegate, and proposal records,
including delegated voting power, proposal timestamps, vote totals, voter
counts, and quorum information. CryptoCompare provides daily USD price
histories and market information used to construct token-size and return
variables. Bitcoin is used as a common benchmark for systematic crypto-market
movements.

The governance and market snapshots were collected on 5 October 2026. Return
and governance-event observations span January 2021 through September 2026.
Governance characteristics such as delegated voting-power concentration and
current delegate structure are therefore primarily cross-sectional snapshots,
whereas token returns are observed historically. Analyses combining these
variables are consequently interpreted as cross-sectional associations rather
than as evidence that the October 2026 governance state caused earlier market
outcomes.

\paragraph{Unit of observation}
The empirical unit is a DAO identified by a fungible governance-token contract
deployed on a mainnet chain. Tally may contain several organizations or governor
contracts associated with the same governance token, for example because a
protocol has migrated between governor versions. Organizations sharing the same
mainnet governance-token contract are therefore consolidated into a single DAO.

When several governors are associated with the same DAO, the primary governor
used to construct delegate-level governance measures is selected according to
the largest amount of delegated voting power. Ties are resolved using the most
recent concluded proposal and then proposal activity. Proposal-event records,
by contrast, are collected across all eligible governors associated with the
DAO token.

\paragraph{Sample construction}
Rather than beginning from a manually selected or popularity-ranked subset, the
pipeline enumerates the Tally organization universe and applies a sequence of
explicit eligibility and data-quality rules. Table~\ref{tab:sample-funnel}
summarizes the resulting sample funnel.

\begin{table}[t]
\centering
\caption{Construction of the empirical DAO sample.}
\label{tab:sample-funnel}
\scriptsize
\setlength{\tabcolsep}{3pt}
\begin{tabular}{p{0.12\columnwidth} p{0.64\columnwidth} r}
\hline
\textbf{Rule} & \textbf{Eligibility criterion} & \textbf{Remaining} \\
\hline
U1   & All organizations returned by Tally & 7,864 \\
U2   & At least one recorded governance proposal & 3,234 \\
U3   & Exclude test, mock, staging, and deprecated organizations & 2,295 \\
U3b  & Governance token deployed on a mainnet chain & 743 \\
U4   & Fungible governance token & 680 \\
U5   & One DAO per unique governance-token contract & 636 \\
U6   & CryptoCompare ticker and project-name match & 80 \\
U6b  & One governance-token contract per matched ticker & 69 \\
U7   & At least one governor uses the identified DAO token & 69 \\
U8   & Token match verified or explicitly adjudicated & \textbf{54} \\
\hline
\end{tabular}
\end{table}

The reduction from 80 ticker-matched organizations to 54 final DAOs reflects contract-level deduplication, governor-token consistency checks, and manual adjudication of ambiguous token matches. Excluded cases and adjudication reasons are recorded in the replication package rather than resolved through silent data modification.

Token matching requires particular care because ticker symbols are not unique
across crypto assets. Candidate CryptoCompare matches are therefore checked
against project identity and, where possible, validated by comparing the
on-chain token supply reported through Tally with the supply reported by
CryptoCompare. A supply ratio between 0.5 and 2 is treated as an automatic
verification criterion. Of the 54 retained DAOs, 42 token matches satisfy this
check directly, while 12 are retained following documented manual adjudication.
The corresponding decisions and reasons are recorded separately rather than
implemented through manual modification of the underlying data.

The final governance universe contains 54 DAOs. Delegate data required for
voting-power concentration are available for 52 of them. Daily USD price data
are available for 46 DAOs, and 45 have at least 12 months of return history.
Twenty-five DAOs are classified as active, defined as having at least one
concluded proposal during the 12 months preceding data collection. Because
individual analyses require different combinations of governance, price,
quorum, and size information, their estimation samples are constructed from
this common universe using complete available cases rather than imposing one
fixed sample size on every specification.

No missing governance value is replaced by cross-sectional mean imputation.
Unavailable or invalid observations remain missing, with the corresponding
reason retained in the data pipeline. This allows the sample used by each
analysis to be reconstructed directly from the underlying eligibility and
measurement rules.

\paragraph{Token-holder count}
Token-holder count is the \texttt{tokenOwnersCount} value returned by Tally
for the governance-token/governor record retained for each consolidated DAO at
the 5 October 2026 snapshot. It is a provider-reported count of token-owning
addresses or accounts, not a count of verified unique persons or organizations.
The measure does not resolve multiple addresses under common control,
distinguish custodial or contract addresses, or identify active voters. We use
it only as a cross-sectional proxy for the breadth of the observed ownership
base.

The complete data pipeline is reproducible from cached source responses and
records both sample exclusions and token-matching decisions. Separate processed
datasets are generated for the DAO cross-section, monthly returns, daily
returns, and governance-event windows. This separation is important because
data availability for cross-sectional governance analysis, return analysis, and
proposal-event analysis differs and the corresponding samples should not be
interpreted as sequential stages of a single estimation sample.

\subsection{Measures of Effective Decentralization}
\label{sec:governance-measures}

Effective decentralization is operationalized through three complementary
dimensions: the distribution of delegated voting power, realized governance
participation, and the breadth of the coalition sufficient to satisfy quorum.
These dimensions are analyzed separately rather than combined into a single
``governance quality'' or decentralization score because they capture distinct
properties of effective decision-making authority.

All delegate-based measures are constructed from the primary governor selected
for each DAO as described in Section~\ref{sec:empirical-data}. Delegate voting
weights and total delegated voting power therefore refer to the same governor
state.

\subsubsection{Voting-Power Concentration}

Let $w_{ij}$ denote the delegated voting power of delegate $j$ in DAO $i$ and
let $W_i$ denote the total delegated voting power reported for the corresponding
governor. Delegate $j$'s voting-power share is
\begin{equation}
    s_{ij}
    =
    \frac{w_{ij}}{W_i}.
    \label{eq:delegate-share}
\end{equation}

The primary concentration measure is the Herfindahl--Hirschman Index (HHI)
computed over the 100 largest delegates:
\begin{equation}
    HHI_i^{(100)}
    =
    \sum_{j=1}^{K_i} s_{ij}^{2},
    \qquad
    K_i = \min(100,N_i),
    \label{eq:hhi}
\end{equation}
where $N_i$ is the number of delegate records available for DAO $i$. Larger
values indicate that delegated voting authority is concentrated among fewer
actors, whereas smaller values indicate a more dispersed distribution.

The data-collection pipeline retrieves up to 500 delegates per governor.
Accordingly, we also compute an HHI using all collected delegates,
$HHI_i^{(\mathrm{all})}$, as a robustness measure. Because both indices use
total delegated voting power as the denominator, the completeness of the
top-100 approximation can be evaluated directly through
\begin{equation}
    Coverage_i^{(100)}
    =
    \sum_{j=1}^{K_i}s_{ij}.
    \label{eq:top100-coverage}
\end{equation}

We complement HHI with cumulative top-$k$ voting-power shares,
\begin{equation}
    C_i(k)
    =
    \sum_{j=1}^{k}s_{ij},
    \qquad
    k \in \{1,5,10\},
    \label{eq:topk-share}
\end{equation}
which provide a more interpretable indication of the voting power controlled
by the largest delegates. We additionally compute the minimum number of the
largest delegates whose cumulative voting power exceeds one half of total
delegated voting power:
\begin{equation}
    NC_i
    =
    \min
    \left\{
        k:
        \sum_{j=1}^{k}s_{ij}\geq 0.5
    \right\},
    \label{eq:nakamoto-governance}
\end{equation}
which we refer to as the governance Nakamoto coefficient. These measures are
used jointly to distinguish concentration around a single dominant delegate
from concentration distributed across several large delegates. They describe
the structural distribution of voting authority and do not imply coordination
or common control among the corresponding addresses.

\subsubsection{Governance Participation}

Participation captures whether formally available governance authority is
actually exercised. We use two complementary measures because address-level
participation and voting-power participation need not coincide.

The first is the \emph{delegate participation rate}. For DAO $i$,
\begin{equation}
    Participation_i
    =
    \frac{\overline{Voters}_i}{Delegates_i},
    \label{eq:participation}
\end{equation}
where $\overline{Voters}_i$ is the mean number of voting addresses across up
to the 20 most recent concluded, non-canceled proposals and $Delegates_i$ is
the current number of registered delegates. If fewer than 20 eligible
proposals are available, all available proposals are used.

This measure captures address-level participation intensity relative to the
recorded delegate population. It should not be interpreted as the fraction of
voting power exercised or as a measure of deliberation quality. Moreover, its
numerator is historical whereas the delegate denominator reflects the current
governance state. When the resulting ratio exceeds one, indicating that the
historical voter count is incompatible with the current delegate denominator,
the observation is treated as unavailable rather than imputed.

To complement this address-based measure, we construct a
\emph{supply-normalized voting turnout} measure,
\begin{equation}
    Turnout_i
    =
    \frac{\overline{VotesCast}_i}{Supply_i},
    \label{eq:voting-turnout}
\end{equation}
where $\overline{VotesCast}_i$ is the mean total voting power cast across the
same recent proposal window and $Supply_i$ is the governance token's total
supply. This measure captures the amount of token voting power mobilized in
governance relative to token supply. It is not equivalent to turnout relative
to currently delegated voting power, but it avoids using the potentially large
and historically inconsistent registered-delegate count as its denominator.
Observations with incompatible voting and token-supply units are treated as
unavailable.

\subsubsection{Effective Quorum Coalition}

Quorum specifies the voting-power threshold that must be reached for a
governance vote to satisfy its formal participation requirement. Quorum alone,
however, does not reveal how many of the largest observed delegate positions
are arithmetically sufficient to reach that threshold.

For each DAO, the quorum value $q_i$ is constructed as the median quorum of up
to the five most recent concluded proposals with usable quorum information.
Proposals recorded as successfully passed despite total votes cast falling
below their reported quorum are treated as data inconsistencies and excluded
from this calculation. When no valid proposal-level quorum is available, the
governor-level quorum parameter is used as a fallback.

Delegates are ordered from largest to smallest voting power. We define the
\emph{Effective Quorum Coalition} as
\begin{equation}
    Q_i
    =
    \min
    \left\{
        k:
        \sum_{j=1}^{k} w_{ij}
        \geq q_i
    \right\}.
    \label{eq:quorum-coalition}
\end{equation}

A smaller $Q_i$ indicates that the quorum threshold can be satisfied by fewer
of the largest delegates, whereas a larger value requires voting power to be
combined across a broader set of actors. We refer to $Q_i$ as the
\emph{Effective Quorum Coalition} (EQC). Importantly, the term ``coalition'' is
used arithmetically: $Q_i$ denotes a sufficient weight set, not an observed or
coordinated group. For example, $Q_i=1$ means that the largest delegate
possesses enough voting power to satisfy quorum alone; it does not imply that
the delegate can unilaterally pass a proposal, since passage may additionally
require majority, supermajority, or other decision conditions. When $Q_i=1$, we
refer to the case as \emph{single-delegate quorum sufficiency} (SDQS). SDQS
identifies DAOs in which the largest delegate alone can satisfy the quorum
threshold, conditional on that delegate actually voting.

The calculation also distinguishes cases in which a valid $Q_i$ cannot be
observed. If the quorum exceeds all currently delegated voting power, the DAO
is classified as having a quorum that is unreachable under the current
delegation state. Cases with missing delegation data, missing quorum
information, incompatible quorum units, or insufficiently observed delegate
power are recorded separately rather than assigned an arbitrary coalition
size. Thus, only DAOs for which the observed delegate distribution reaches a
valid quorum receive an exact value of $Q_i$.

The data do not provide person- or entity-level resolution across delegate
records. Separate delegate positions may therefore be controlled by the same
person or organization, while one entity may also operate several addresses.
Accordingly, $Q_i$ counts observed delegate positions and should not be read as
a count of independently controlled actors.

As a secondary robustness check, we additionally calculate the \emph{Quorum Coalition Ratio},
\begin{equation}
    Q_i^{ratio}
    =
    \frac{Q_i}{Delegates_i}.
    \label{eq:quorum-normalized}
\end{equation}

This normalized measure is retained only as a secondary robustness
specification. Because the registered delegate count can contain inactive
addresses and is itself related to token-market scale, normalization may
introduce mechanical relationships that are absent from the raw coalition
measure. The raw Effective Quorum Coalition $Q_i$ is therefore used as the
primary measure of quorum-coalition breadth.

\subsection{Market Variables}
\label{sec:market-variables}

Token-market variables are constructed from CryptoCompare daily USD price
histories. Daily log returns are calculated on consecutive calendar days,
while monthly returns are calculated from month-end closing prices. For token
$i$ at time $t$,
\begin{equation}
    r_{it}
    =
    \ln(P_{it})
    -
    \ln(P_{i,t-1}).
    \label{eq:log-return}
\end{equation}

Bitcoin is used as a common benchmark for systematic crypto-market movements.
The corresponding BTC-relative return is
\begin{equation}
    ER_{it}
    =
    r_{it}
    -
    r_{\mathrm{BTC},t}.
    \label{eq:excess-return}
\end{equation}

The term BTC-relative return is preferred to fully risk-adjusted abnormal
return because Bitcoin represents a parsimonious market benchmark rather than
a complete crypto-asset pricing model.

Token-market scale is measured using token-based valuation variables. The primary
measure is live-token fully diluted valuation (FDV), computed from token price and supply for tokens with observable trading activity near the end of the sample. For DAO $i$,
\begin{equation}
    FDV^{live}_i
    =
    \overline{P}_{i,\mathrm{Sep\,2026}}
    \times
    Supply_i,
    \label{eq:fdv-live}
\end{equation}
where $\overline{P}_{i,\mathrm{Sep\,2026}}$ is the mean token price during
September 2026 and $Supply_i$ is the governance-token supply retained by the
matching pipeline following automated verification or documented
adjudication.
$FDV^{live}_i$ is defined only for tokens with positive trading volume during
that month, thereby avoiding the use of carried-forward prices for inactive
or effectively non-traded tokens.

Two alternative token-market-scale measures are used for robustness. The first is the
CryptoCompare market-capitalization snapshot collected on 5 October 2026,
after excluding unavailable, stale, or supply-inconsistent observations. The
second is a trading-life FDV measure constructed from the mean observed token
price on days with positive trading volume multiplied by token supply. These
alternative measures allow the analysis to assess whether conclusions depend
on a particular representation of token-market scale.

\subsection{Statistical Analysis}
\label{sec:statistical-analysis}

The quantitative analysis addresses four complementary questions. First, we
test whether observable governance characteristics are associated with
token-market scale. Second, we examine whether the same characteristics are associated with
average token returns. Third, we quantify the contribution of systematic
crypto-market exposure to monthly DAO-token returns. Fourth, we test whether
returns around governance proposal completion differ from comparable non-event
periods.

All cross-sectional models use complete available cases for the variables
required by a given specification. Statistical inference is complemented by
non-parametric and resampling procedures where appropriate because several
governance and market variables are strongly skewed and the effective sample
sizes are modest.

\subsubsection{Governance and Token-Market Scale}

The primary token-market-scale outcome is the logarithm of live-token FDV,

\begin{equation}
    Y_i
    =
    \ln(FDV^{live}_i).
    \label{eq:dao-size-outcome}
\end{equation}

CryptoCompare market capitalization and trading-life-average FDV are analyzed
as alternative token-market-scale measures.

The analysis first estimates Spearman rank correlations between token-market scale and
governance characteristics that do not mechanically divide by the number of
delegates. These include HHI, Top-1 and Top-10 voting-power shares, the
governance Nakamoto coefficient, the raw Effective Quorum Coalition $Q_i$, and
supply-normalized voting turnout. Delegate and token-holder counts are also
examined as measures of the breadth of the governance and ownership bases.
Confidence intervals are reported for the rank correlations together with the
minimum correlation detectable with 80\% power at the corresponding sample
size.

The regression analysis then uses nested specifications to distinguish
voting-power concentration from the scale of the delegate population. The
basic model is

\begin{equation}
    \ln(Size_i)
    =
    \alpha
    +
    \beta_1 \ln(HHI_i)
    +
    \varepsilon_i,
    \label{eq:size-m1}
\end{equation}
followed by
\begin{equation}
    \ln(Size_i)
    =
    \alpha
    +
    \beta_1 \ln(HHI_i)
    +
    \beta_2 \ln(Delegates_i)
    +
    \varepsilon_i.
    \label{eq:size-m2}
\end{equation}

A further specification introduces supply-normalized voting turnout,
\begin{equation}
\begin{aligned}
    \ln(Size_i) ={}& \alpha + \beta_1 \ln(HHI_i)
    + \beta_2 \ln(Delegates_i) \\
    &+ \beta_3 \ln(Turnout_i) + \varepsilon_i,
\end{aligned}
    \label{eq:size-m3}
\end{equation}
while an additional robustness model controls for token age and an indicator
for deployment on Ethereum mainnet.

OLS inference uses HC3 heteroskedasticity-robust standard errors. For the
bivariate HHI--size relationship, uncertainty is additionally assessed using
a 1,000-replication pairs bootstrap with a fixed random seed.

The delegate participation rate and Quorum Coalition Ratio are retained as
secondary comparisons with the earlier empirical specification. Both divide
by the registered delegate count, which is itself strongly related to
token-market scale, and are therefore not treated as primary evidence about
the relationship between decentralization and token-market scale.

The analysis is explicitly associational. Neither the cross-sectional
regressions nor the rank correlations establish whether governance structure
affects token-market scale, whether higher-valued DAO tokens are associated
with different governance structures, or whether both are jointly determined
by organizational maturity and other characteristics.

\subsubsection{Governance and Token Returns}

Cross-sectional return models examine whether static governance
characteristics contain information about average token performance. The
analysis is restricted to tokens with at least 12 months of observed monthly
returns.

For DAO $i$, let $\overline{r}_i$ denote its average monthly log return. The
principal specification is
\begin{equation}
\begin{aligned}
    \overline{r}_i ={}& \alpha + \beta_1 \ln(HHI_i)
    + \beta_2 \ln(Participation_i) \\
    &+ \beta_3 \ln(Q_i^{ratio}) + \varepsilon_i,
\end{aligned}
    \label{eq:return-crosssection}
\end{equation}
with nested models introduced sequentially to avoid forcing the smallest
complete-case sample on every specification. The same models are estimated
using average BTC-relative monthly returns,
\begin{equation}
    \overline{ER}_i
    =
    \overline{r}_i
    -
    \overline{r}_{\mathrm{BTC},i},
\end{equation}
to reduce the influence of common crypto-market movements.

For comparability with the preceding token-market-scale analysis, non-parametric
Spearman correlations are also reported. OLS models use HC3 robust standard
errors.

A specification that additionally controls for end-of-sample token-market scale is
treated only as a sensitivity analysis. Because the size measure is observed
at the end of a return history extending back to 2021, conditioning historical
returns on that later valuation can introduce a mechanical temporal
relationship. The principal return conclusions therefore do not rely on this
token-market-scale-controlled model.

\subsubsection{Systematic Crypto-Market Exposure}

To quantify systematic crypto-market exposure, monthly DAO-token returns are
first regressed on contemporaneous Bitcoin returns:
\begin{equation}
    r_{it}
    =
    \alpha
    +
    \beta_{\mathrm{BTC}} r_{\mathrm{BTC},t}
    +
    \varepsilon_{it}.
    \label{eq:btc-baseline}
\end{equation}

The governance analysis is then estimated on a common sample using nested
models. The principal governance specification is
\begin{equation}
\begin{aligned}
    r_{it} ={}& \alpha + \beta_{\mathrm{BTC}} r_{\mathrm{BTC},t}
    + \beta_1 \ln(HHI_i) \\
    &+ \beta_2 \ln(Delegates_i) + \varepsilon_{it},
\end{aligned}
    \label{eq:btc-governance}
\end{equation}
followed by
\begin{equation}
\begin{aligned}
    r_{it} ={}& \alpha + \beta_{\mathrm{BTC}} r_{\mathrm{BTC},t}
    + \beta_1 \ln(HHI_i) \\
    &+ \beta_2 \ln(Delegates_i)
    + \beta_3 \ln(FDV^{live}_i) \\
    &+ \varepsilon_{it}.
\end{aligned}
    \label{eq:btc-governance-size}
\end{equation}

Using a common estimation sample across the nested models permits direct
comparison of their explanatory power. A specification using the delegate
participation rate and Quorum Coalition Ratio is retained as a secondary
comparison with the earlier analysis.

Because each DAO contributes repeated monthly observations, standard errors are
clustered at the DAO level and inference uses the corresponding cluster-based
$t$ distribution, following established guidance for within-cluster dependence
\cite{cameron2015cluster}. Governance and size coefficients are additionally
evaluated using a restricted wild-cluster bootstrap with 999 replications,
which is particularly useful when the number and sizes of DAO clusters are
limited \cite{mackinnon2017wild}.

The governance variables are cross-sectional snapshots and therefore do not
vary across months within a DAO. The panel should consequently be interpreted
primarily as an analysis of contemporaneous crypto-market exposure and
cross-sectional governance differences, not as evidence about the effect of
within-DAO governance changes over time.

\subsubsection{Governance Proposal Event Study}

The proposal-event analysis follows the standard logic of measuring abnormal
returns in a bounded window around an identifiable event
\cite{brown1985event,mackinlay1997event}. It examines token returns around the
end of voting on concluded governance proposals. Proposal completion is used
because it provides a consistently observable timestamp across organizations.
It is not assumed to represent the unique moment of information arrival:
proposal content, voting trajectories, delegate positions, and likely outcomes
may become observable before voting closes.

For proposal $e$ of DAO $i$, the BTC-relative daily return is
\begin{equation}
    AR_{i,e,t}
    =
    r_{i,t}
    -
    r_{\mathrm{BTC},t},
    \label{eq:event-ar}
\end{equation}
and cumulative BTC-relative return over the symmetric $[-5,+5]$ day window is
\begin{equation}
    CAR_{i,e}[-5,+5]
    =
    \sum_{k=-5}^{+5}
    AR_{i,e,\tau_e+k},
    \label{eq:event-car}
\end{equation}
where $\tau_e$ denotes the proposal end date. Only proposals with a complete
11-day price window are retained. The resulting event dataset contains 1,806
proposals from 35 DAOs, corresponding to 1,391 distinct DAO--event dates over
January 2021--September 2026.

A raw negative CAR is not sufficient to identify a governance-specific price
effect because a DAO token may systematically underperform Bitcoin outside
governance windows. We therefore construct a DAO-specific non-event baseline.
For DAO $i$, let
\begin{equation}
    B_i
    =
    \frac{1}{M_i}
    \sum_{m=1}^{M_i}
    CAR^{non-event}_{im},
    \label{eq:event-baseline}
\end{equation}
where the baseline is the mean BTC-relative CAR across all complete 11-day
windows for that DAO containing no governance-event day. The event-specific
excess CAR is then
\begin{equation}
    ECAR_{i,e}
    =
    CAR_{i,e}
    -
    B_i.
    \label{eq:excess-car}
\end{equation}

This within-DAO comparison reduces sensitivity to persistent differences in
BTC-relative performance across tokens.

Cross-sectional correlation and clustered event dates can distort conventional
event-study tests \cite{kolari2010event}. Dependence is therefore addressed
using three event samples. The first retains all
eligible proposals. The second collapses multiple proposals from the same DAO
that end on the same date into a single event date. The third constructs a
non-overlapping sample in which retained event dates for a given DAO are at
least 11 days apart. Mean CAR and excess CAR are evaluated using standard
errors clustered by DAO, two-way clustering by DAO and calendar month, and a
DAO-level test in which each DAO contributes one average observation.

As a complementary robustness check, placebo event windows are sampled from
each DAO's event-free 11-day windows while preserving the number of events
contributed by that DAO. The procedure uses 1,000 replications with a fixed
random seed and reports both proposal-weighted and equal-DAO-weighted placebo
distributions. Because randomized windows do not by themselves account for
common calendar clustering, placebo inference is treated as complementary to,
rather than a substitute for, clustered and non-overlapping event
specifications.

Finally, mean BTC-relative returns on DAO-days falling inside any event window
are compared with returns on the remaining observed days. Together, these
tests distinguish a raw return pattern around governance proposals from a
return pattern that is unusual relative to the same tokens' non-event
behavior.

\subsection{Missing Data and Robustness}
\label{sec:missing-robustness}

No missing governance or market observation is replaced through mean
imputation. Each statistical specification is estimated on complete available
cases for the variables it requires, while the reason for unavailable or
invalid observations is retained during data construction. Consequently,
sample sizes may differ across governance measures and statistical models.

Robustness analysis is organized around the principal sources of measurement
and specification uncertainty. Voting-power concentration is examined using
top-100 HHI, HHI based on all collected delegates, Top-1 and Top-10 shares,
top-100 coverage, and the governance Nakamoto coefficient. Quorum structure is evaluated using the raw EQC defined in Section~\ref{sec:governance-measures}. The normalized Quorum Coalition Ratio is retained only as a secondary robustness specification, because normalization by the registered delegate count can introduce mechanical relationships with token-market scale.

The token-market-scale analysis compares live-token FDV with CryptoCompare market
capitalization and trading-life-average FDV, reports confidence intervals and
power for rank correlations, and uses bootstrap inference for the principal
HHI specification. Results are also examined separately among recently active
DAOs.

For the monthly return panel, DAO-clustered inference is supplemented with
wild-cluster bootstrap tests. For governance proposal events, robustness is
assessed through distinct event dates, non-overlapping within-DAO event
windows, DAO- and DAO--month-clustered inference, DAO-level aggregation, and
1,000-replication placebo tests. These procedures are intended to prevent
conclusions from depending on a single concentration measure, sample
definition, weighting convention, or independence assumption.

\subsection{Validity and Reproducibility}
\label{sec:validity-reproducibility}

The combined research design has distinct validity constraints for its
systematization and empirical components. Governance mechanisms evolve over
time, protocol documentation may lag deployed contracts, and not all
governance activity is represented through a common platform. The empirical
sample is specifically conditioned on on-chain governance indexed by Tally and
on the availability of corresponding CryptoCompare token-market information;
it therefore does not represent the complete population of DAOs and excludes
governance conducted exclusively through off-chain systems such as Snapshot.

A second limitation concerns temporal alignment. Delegate distributions,
participation denominators, quorum structure, and other governance variables
are primarily measured from the governance state observed in October 2026,
whereas return and proposal-event histories extend from January 2021 through
September 2026. Historical analyses involving the static governance variables
are therefore interpreted as associations between current governance structure
and observed historical outcomes, not as causal estimates of the effect of
governance at the time those returns were realized.

Several data-quality risks are addressed directly in the construction
pipeline. Ticker-based market-data matches are validated against project
identity and token supply; duplicate and test governance organizations are
removed; primary governors are selected according to explicit rules; stale or
inconsistent market observations remain unavailable rather than being
imputed; and quorum values with incompatible or implausible units are recorded
separately. These procedures reduce, but cannot eliminate, measurement error
arising from changing governance implementations, token migrations, sparse
trading, and platform-specific data coverage.

The empirical pipeline is fully script-based. Raw Tally and CryptoCompare
responses are cached with collection timestamps, exclusions and token-matching
decisions are recorded explicitly, and the pipeline regenerates the cleaned
DAO cross-section, daily and monthly return datasets, proposal-event windows,
statistical tables, figures, and LaTeX result macros. This separation between
raw inputs, documented decisions, processed datasets, and generated manuscript
outputs reduces the risk of manual inconsistencies when the analysis is rerun.

For the SoK component, reproducibility is supported by the explicit search
vocabulary, eligibility rules, source hierarchy, case-selection criteria, and
coding dimensions reported above. The replication package preserves the
empirical scripts, cached source responses, exclusion log, token-matching
decisions, processed tables, and figure inputs. Protocol documentation remains
version-sensitive; the cited source and observation date therefore form part
of each qualitative claim's scope.

\section{DAO Governance Design Space}
\label{sec:design-space}

The comparative evidence can be organized as a sequence of authority allocations rather than as a list of isolated mechanisms. We use six recurring governance layers to structure this analysis: agenda access, voting and delegation, decision thresholds, execution constraints, emergency authority, and external dependencies. At each layer, the relevant question is not merely which mechanism is used, but how that mechanism allocates, concentrates, exercises, or constrains authority. Thus, a governance system must determine who may place an issue on the binding agenda, how preferences are represented and aggregated, which minimum weight sets satisfy decision thresholds, how an approved action becomes executable, who may intervene when the ordinary process is too slow, and which external feeds or infrastructure dependencies can trigger exceptional governance action.

\subsection{Governance Lifecycle}
\label{sec:governance-lifecycle}

Mature DAOs commonly separate social deliberation from binding execution
\cite{reijers2021code,schadler2023decisionmaking,ozGovernanceGuide}. The precise implementation varies,
but the recurring sequence is discussion, proposal formation, binding voting,
threshold evaluation, queued execution, and state change. Table~\ref{tab:lifecycle}
shows the function of each stage and the authority question it creates.

\begin{table*}[t]
\centering
\caption{Governance lifecycle and corresponding authority questions.}
\label{tab:lifecycle}
\small
\setlength{\tabcolsep}{5pt}
\begin{tabular}{p{0.14\textwidth}p{0.31\textwidth}p{0.43\textwidth}}
\hline
\textbf{Stage} & \textbf{Governance function} & \textbf{Authority question} \\
\hline
Deliberation & Develop and contest an idea in forums or community channels &
Who can obtain attention, frame the issue, and shape the feasible proposal set? \\
Proposal formation & Translate an idea into a formally admissible action &
Who can propose, sponsor, or certify that procedural conditions are met? \\
Voting & Record preferences directly or through delegates &
How is voting power assigned, delegated, and made accountable? \\
Decision & Apply quorum, majority, supermajority, or vote-differential rules &
How many observed delegate positions are sufficient to satisfy a decision threshold? \\
Queue and review & Delay an authorized action before it changes protocol state &
Who can inspect, cancel, veto, or accelerate a queued action? \\
Execution and response & Execute the action or invoke exceptional controls &
Who controls the final transaction, and who may pause or override operation under stress? \\
\hline
\end{tabular}
\end{table*}

This staged structure creates multiple, potentially different loci of control.
A system may be open at the discussion stage but restrictive at proposal
creation; broadly eligible to vote but concentrated in delegated weight; or
collectively governed under ordinary conditions but dependent on a small
council during an incident. Decentralization at one stage therefore cannot be
assumed to extend to the entire lifecycle.

\subsection{Agenda Access and Proposal Formation}
\label{sec:agenda-access}

Agenda access determines which preferences can become binding choices. Common
designs include permissionless submission, minimum voting-power thresholds,
delegate sponsorship, staged temperature checks, and process-mediated intake.
High proposal thresholds filter spam and force proponents to demonstrate
support before consuming governance attention, but they also make agenda
access dependent on wealthy token holders or established delegates. Lower
thresholds broaden access while increasing review and coordination costs.

Off-chain forums and signaling votes soften this trade-off by allowing ideas
to be tested before an executable proposal is created. Deployed processes in
Uniswap, Aave, and Compound illustrate how forum deliberation, preliminary
checks, and formal on-chain submission can be combined in different sequences
\cite{uniswapProcess,aaveGovProcessV1,compoundGovernanceDocs}. These stages do
not eliminate gatekeeping: moderators, facilitators, recognized delegates, and
established community norms can still affect which issues progress. Agenda decentralization must therefore be evaluated using both the formal proposer rule and the practical path from discussion to a binding payload. In self-funding governance systems, this path is especially consequential because proposal rights are coupled to control over budget withdrawals and ongoing financing \cite{Chistiakov2020}.

\subsection{Voting, Delegation, and Representation}
\label{sec:voting-delegation-design}

Token-weighted voting remains the dominant aggregation mechanism in the
examined systems, frequently supplemented by delegation
\cite{sun2024makerdao,weidener2025delegated}. Protocol guidance for Uniswap and
Aave documents direct or delegated voting workflows, while ENS has introduced
tools for allocating delegation across multiple representatives
\cite{uniswapVotingGuide,aaveVoting,ensMDM}. Delegation reduces the cost of
continuous participation and permits specialization: token holders can assign
their votes to actors who monitor proposals and publish rationales. It also
creates a representative layer whose distribution may differ sharply from the
underlying token-holder population.

This dual effect explains why delegate count and delegate concentration must be
measured separately. A DAO may have thousands of registered delegates while a
small subset holds most delegated voting power. Conversely, a relatively small
delegate population may be less concentrated if voting weight is evenly
distributed. Transparency, rationale publication, conflict disclosure, and
low-friction re-delegation can make representative authority more contestable,
but none mechanically prevents concentration.

Participation incentives add another design layer. Compensation can improve
continuity and professionalize review, while performance criteria can make
delegates more accountable. Poorly designed programs may entrench incumbents
or reward easily measured activity rather than independent judgment. The
relevant question is not simply whether participation is paid, but which actors
qualify, which behavior is rewarded, and whether delegation remains open to
challenge.

\subsection{Decision Thresholds and Coalition Requirements}
\label{sec:decision-rules}

Quorum and passage rules translate voting weights into a collective decision;
their concrete implementation is visible in governor contracts and protocol
documentation \cite{ozGovernanceApi,compoundGovernanceDocs}.
Quorum protects against decisions made under very low participation, while
majority, supermajority, or vote-differential rules determine whether the
proposal passes. These rules are often interpreted in isolation, but their
practical effect depends on the distribution of voting power.

A numerically high quorum can still be satisfied by one highly delegated actor,
whereas a lower threshold may require a broad coalition in a dispersed
distribution. Conversely, a quorum above currently delegated voting power can
make governance inoperative even though the rule appears conservative. The Effective Quorum Coalition operationalized in Section~\ref{sec:governance-measures} and estimated in Section~\ref{sec:effective-quorum}
captures this interaction. It measures arithmetic sufficiency for quorum, not
the complete coalition required for proposal passage, and therefore complements
rather than replaces protocol-specific decision rules.

\subsection{Execution Constraints}
\label{sec:execution-constraints}

Approval by token holders does not necessarily produce an immediate
state-changing action. Many DAO governance architectures separate
\emph{authorization} from \emph{execution} through queues, timelocks, staged
execution procedures, or other post-vote constraints. These mechanisms create
an interval in which an approved action is known but has not yet modified the
underlying protocol state.

This separation serves an important governance function. Voting determines
whether an action has received sufficient collective authorization, whereas
execution constraints determine when and under what conditions that
authorization becomes operational. A proposal may therefore satisfy quorum and
its applicable passage rule while remaining subject to additional delay,
review, or execution requirements before it can affect protocol contracts,
treasury assets, or system parameters.

\paragraph{Timelocks and execution delays}

A timelock requires an approved governance action to remain queued for a
predefined interval before execution. During this period, the proposed state
transition is publicly observable but cannot yet take effect
\cite{ozTimelock,ozGovernanceApi}. The delay provides participants, developers,
security teams, and other monitoring actors with an opportunity to inspect the
authorized transaction and identify unexpected or malicious behavior before
execution.

Timelocks therefore create a temporal separation between collective approval
and irreversible or costly state change. This distinction is especially
important for contract upgrades, parameter changes, treasury transfers, and
other actions whose consequences may be difficult to reverse after execution.

The protection provided by a delay nevertheless depends on what can happen
during the review window. A timelock alone does not prevent an unsafe action
from eventually executing. Its practical value is greatest when participants
can detect the problem and some authorized mechanism exists to cancel, veto,
pause, or otherwise prevent execution before the delay expires.

\paragraph{Review windows and queued actions}

The interval between approval and execution can therefore be understood as a
\emph{review window}. During this period, the governance community can verify
that the executable payload corresponds to the approved proposal, examine its
technical consequences, and monitor for implementation errors or newly
identified security concerns.

The existence of a review window changes the temporal structure of governance
authority. Before the vote, participants determine whether an action should be
authorized. After approval but before execution, attention shifts toward
verification and intervention. The relevant question is no longer only
\emph{whether the proposal was approved}, but also \emph{who can prevent or
modify its execution if a problem is identified}.

This creates an important connection between ordinary execution constraints
and emergency authority. A timelock can provide time for intervention, but a
separate guardian, council, veto mechanism, or governance action may be required
to make use of that time. Execution delay and emergency authority should
therefore be evaluated as complementary rather than interchangeable safeguards.

\paragraph{Upgrade execution}

Execution constraints are particularly consequential for protocol upgrades.
An approved upgrade may replace implementation code, modify administrative
permissions, alter risk parameters, or change the governance system itself.
Because such actions can affect the future distribution of authority as well as
the protocol's technical behavior, the execution pathway becomes part of the
governance design rather than a purely operational detail.

Different architectures can place execution authority in different locations.
Some governance systems permit approved actions to become executable directly
through a governor and timelock, whereas others involve additional executor,
multisignature, or administrative components. The relevant decentralization
question is therefore not only who approves an upgrade, but whether the same
actors also control its final execution and whether another privileged actor can
delay, veto, modify, or bypass that execution path.

\paragraph{Execution failure surfaces}

Execution constraints mitigate some governance risks while introducing others.
A very short or absent delay can reduce the time available to detect a harmful
proposal after approval. A long delay provides more opportunity for review but
can slow legitimate upgrades and corrective actions. Additional executors,
vetoes, or cancellation mechanisms may improve safety while also introducing
privileged control points whose authority must itself be governed.

The principal trade-off is therefore:

\begin{quote}
\textbf{Responsiveness versus execution safety.}
\end{quote}

Shorter execution paths increase governance responsiveness but reduce the time
available for verification and intervention. Longer or more constrained
execution paths increase the opportunity for review, but they can delay
legitimate state changes and become problematic when the protocol must react
quickly to an operational or security incident.

For the comparative framework, execution constraints are therefore evaluated
along several dimensions: whether approved actions are queued; the duration
and configurability of execution delays; whether queued actions can be canceled
or vetoed; the actor or contract that ultimately performs execution; and
whether upgrades, treasury actions, and routine parameter changes use
different execution paths. The framework deliberately records mechanism and
authority rather than treating a longer delay as universally safer: the value
of a review window depends on monitoring and on the existence of a legitimate
intervention path.

\subsection{Exceptional Authority and External Dependencies}
\label{sec:exceptional-authority-design}

Ordinary token-holder governance is often supplemented by guardians, security councils, or multisignature committees that can pause selected operations, cancel queued actions, or coordinate urgent upgrades
\cite{aaveGovDocsGuardian,arbitrumSecurityCouncil,makerGovModule,makerwhitepaper2020}. These actors
reduce response latency but create a second locus of authority. Their governance
quality depends on scope, signer structure, activation conditions, duration,
revocability, public auditability, and whether ordinary governance can replace
the actor or review the intervention.

External systems can create comparable authority dependencies. Price and event
oracles, bridges, and cross-domain messaging systems determine inputs that
smart contracts cannot verify from local state alone
\cite{eskandari2021sokOracles,chainlinkDataFeeds}. Governance therefore extends
to provider selection, update and deviation parameters, fallback feeds,
dispute pathways, pause authority, and replacement procedures. An oracle can
be technically decentralized while its selection or emergency replacement is
controlled by a narrow set of actors; conversely, a centralized data source may
be surrounded by transparent limits and rapid governance replacement. Oracle
governance must consequently be evaluated as part of the protocol's authority
configuration rather than as an independent infrastructure choice.

\subsection{Recurring Design Trade-offs}
\label{sec:design-tradeoffs}

Across the qualitative cases, six trade-offs recur. Lower proposal barriers
increase openness but raise screening costs. Delegation increases continuity
but can concentrate representative authority. Higher quorums can protect
legitimacy but can also create deadlock and may remain weak when voting power is
concentrated. Longer execution delays improve reviewability while slowing
remediation. Emergency authority improves response speed while
creating privileged control. Finally, external dependencies such as oracles and
bridges can extend protocol functionality while introducing failure or control
points that may require governance intervention. None of these mechanisms is intrinsically
decentralizing or centralizing; its effect depends on the authority it moves,
the conditions under which that authority can be exercised, and the safeguards
that constrain it.

\section{Measuring Effective Decentralization}
\label{sec:effective-decentralization}

The governance-design analysis identifies where authority is formally allocated,
but formal architecture does not establish how that authority is distributed or
exercised in practice. We therefore examine three complementary dimensions of
effective decentralization: voting-power concentration, governance
participation, and the breadth of the coalition sufficient to satisfy quorum.
The construction of these measures is described in
Section~\ref{sec:governance-measures}; this section reports their empirical
distribution across the final DAO sample.

\subsection{Voting-Power Concentration}
\label{sec:voting-power-concentration}

Delegate-level voting data are available for 52 of the 54 DAOs. The distribution
is characterized by substantial concentration, although the degree of
concentration varies considerably across organizations.

The top-100 HHI has a mean of 0.333 and a median of 0.252, with observed values
ranging from 0.024 to 1.000. The corresponding HHI calculated using all
collected delegates is virtually identical. This similarity reflects the high
coverage of the top-100 delegate set: mean coverage is 98.9\% and median
coverage is approximately 100\%. Even in the lowest-coverage observation, the
top 100 account for approximately 87.3\% of total delegated voting power.

The cumulative voting shares provide a more direct interpretation of this
concentration. The largest delegate controls, on average, 45.0\% of delegated
voting power, with a median share of 41.6\%. The five largest delegates jointly
control a median of 88.7\%, while the ten largest control a median of 97.8\%.
The governance Nakamoto coefficient---the minimum number of the largest
delegates required to exceed 50\% of delegated voting power---has a median of
2 and a mean of 3.10.

These statistics indicate that nominally large delegate populations can coexist
with voting authority concentrated among a relatively small number of highly
weighted actors. They should not, however, be interpreted using conventional
industrial-organization HHI thresholds or as evidence that large delegates
coordinate. The measures describe the distribution of structural voting
capacity rather than observed coalition behavior.

\subsection{Governance Participation}
\label{sec:participation}

Participation provides a separate perspective on effective decentralization
because the existence of voting rights does not imply that those rights are
actively exercised. Usable delegate-participation observations are available
for 50 DAOs.

The delegate participation rate has a mean of 0.083 and a median of 0.032.
Thus, for the median DAO, the average number of voting addresses observed on
recent proposals corresponds to approximately 3.2\% of the current registered
delegate population. The distribution is strongly right-skewed, with observed
values ranging from zero to approximately 0.589.

Supply-normalized voting turnout produces a similar qualitative picture but
captures voting activity in voting-power rather than address terms. Its
DAO-level mean is 0.094 and cross-DAO median is 0.043, indicating that, for the
median DAO, mean recent-proposal turnout corresponds to approximately 4.3\% of
governance-token supply.

The two measures should not be interpreted interchangeably. Delegate
participation compares voter addresses with the recorded delegate population,
whereas supply-normalized turnout compares votes cast with token supply.
Consequently, a DAO can exhibit limited address-level participation while still
mobilizing substantial voting power if the participating delegates hold large
voting positions. Neither measure, by itself, captures deliberation quality,
delegate independence, or representational legitimacy.

\subsection{Effective Quorum Coalition}
\label{sec:effective-quorum}

As defined in Section~\ref{sec:governance-measures}, a valid EQC can be calculated for 38 of the 54 DAOs.
Among these organizations, the mean number of largest delegates required to
satisfy quorum is 2.71 and the median is 1, with observed values ranging from 1 to 18.

Most strikingly, the largest delegate alone possesses sufficient voting power
to satisfy the quorum threshold in 24 of the 38 DAOs with a valid EQC,
corresponding to approximately 63\% of the observed cases. This
pattern is not confined to inactive organizations: among recently active DAOs
with a valid EQC, 14 of 23, or approximately 61\%, exhibit SDQS.

This result should be interpreted narrowly. As noted above, $Q_i=1$ denotes
arithmetic quorum sufficiency, not unilateral proposal passage: a proposal may
still require a majority, supermajority, vote differential, or other decision
condition. We therefore describe this property as SDQS rather than quorum
capture.

The remaining DAO states are also informative. In 11 cases, the recorded quorum
exceeds the total voting power currently delegated, implying that quorum is
unreachable under the observed delegation state without additional delegation
or another change in voting power. Two further cases have implausible quorum
units, two lack usable delegation data, and one has no usable quorum value.
These observations are retained as distinct governance states rather than being
assigned artificial coalition sizes.

\subsection{Cross-DAO Heterogeneity}
\label{sec:governance-heterogeneity}

Table~\ref{tab:governance-descriptive} summarizes the principal governance
measures. The results reinforce the distinction between concentration,
participation, and quorum-coalition breadth. These dimensions describe related
but non-equivalent properties of governance and need not move together.

\begin{table*}[t]
\centering
\caption{Descriptive statistics for effective-decentralization measures.}
\label{tab:governance-descriptive}
\small
\begin{tabular}{lrrrrrr}
\hline
\textbf{Metric} &
\textbf{$N$} &
\textbf{Mean} &
\textbf{Median} &
\textbf{SD} &
\textbf{Min} &
\textbf{Max} \\
\hline

HHI (top 100) &
52 & 0.333 & 0.252 & 0.274 & 0.024 & 1.000 \\

Top-1 voting share &
52 & 0.450 & 0.416 & 0.271 & 0.047 & 1.000 \\

Top-5 voting share &
52 & 0.796 & 0.887 & 0.232 & 0.222 & 1.000 \\

Top-10 voting share &
52 & 0.883 & 0.978 & 0.176 & 0.393 & 1.000 \\

Top-100 coverage &
52 & 0.989 & 1.000 & 0.028 & 0.873 & 1.003 \\

Governance Nakamoto coefficient &
52 & 3.10 & 2.00 & 3.17 & 1 & 15 \\

Delegate participation rate &
50 & 0.083 & 0.032 & 0.123 & 0.000 & 0.589 \\

Supply-normalized voting turnout &
50 & 0.094 & 0.043 & 0.172 & 0.000 & 1.000 \\

Effective Quorum Coalition &
38 & 2.71 & 1.00 & 3.75 & 1 & 18 \\

\hline
\end{tabular}

\vspace{2pt}
\footnotesize
\parbox{\textwidth}{\textit{Note:} HHI and top-$k$ shares refer to delegated voting power.
The governance Nakamoto coefficient is the minimum number of largest delegates
whose cumulative voting power exceeds 50\% of total delegated voting power.
Effective Quorum Coalition statistics include only DAOs for which the observed
delegation distribution reaches a valid quorum. Top-100 coverage values
marginally above one can arise from small differences between aggregate
delegated-vote totals and the sum of delegate records returned by the API.}
\end{table*}

\begin{figure*}[t]
    \centering
    \includegraphics[width=\textwidth]
    {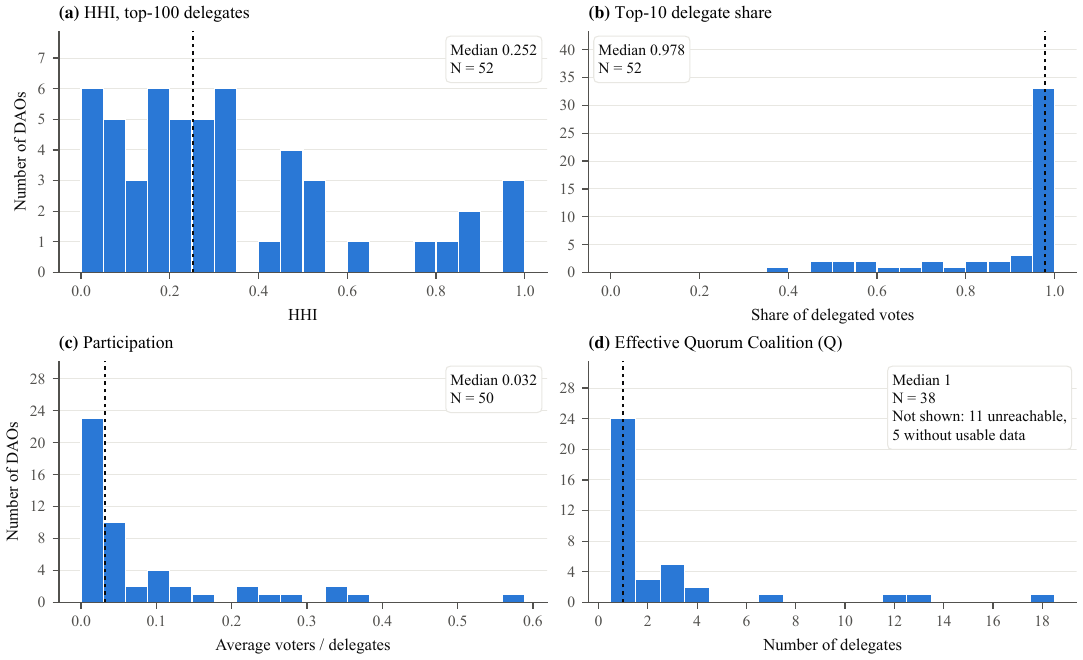}
    \caption{Distribution of effective-decentralization measures.
    Panel~(a) reports HHI based on the 100 largest delegates;
    Panel~(b) reports the cumulative voting-power share of the ten largest
    delegates; Panel~(c) reports the delegate participation rate; and
    Panel~(d) reports the Effective Quorum Coalition for DAOs with a reachable
    quorum. Dashed vertical lines indicate sample medians.}
    \label{fig:governance-distributions}
\end{figure*}

Figure~\ref{fig:governance-distributions} illustrates four aspects of this
heterogeneity. Panel~(a) reports the distribution of voting-power concentration
using HHI; Panel~(b) shows the cumulative voting share of the ten largest
delegates; Panel~(c) reports the delegate participation rate; and Panel~(d)
shows the Effective Quorum Coalition for DAOs with a reachable quorum.

\begin{figure*}[t]
    \centering
    \includegraphics[width=\textwidth]
    {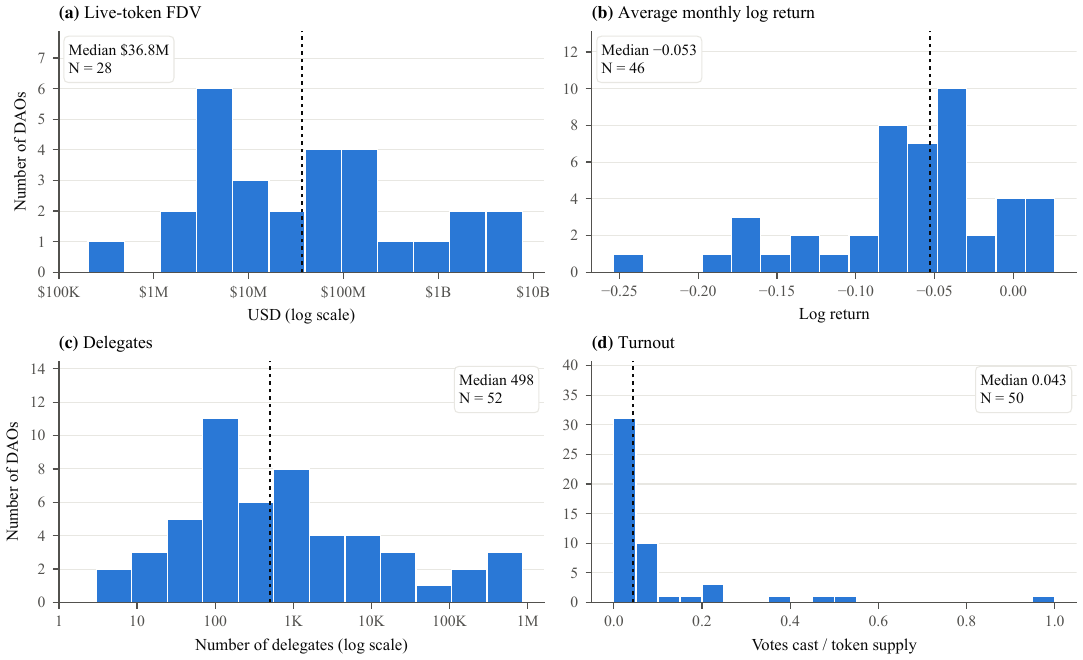}
    \caption{Distribution of token-market and participation variables:
    (a) live-token FDV (log scale); (b) average monthly log return;
    (c) number of registered delegates (log scale); (d) turnout, measured as
    votes cast relative to token supply. Dashed vertical lines indicate sample
    medians.}
    \label{fig:market-distributions}
\end{figure*}

Figure~\ref{fig:market-distributions} complements these governance measures
with the distributions of live-token FDV, average monthly log returns, the
number of registered delegates, and turnout.

Taken together, the descriptive evidence shows why effective decentralization
cannot be inferred from nominal delegate counts or a single governance metric.
The observed DAOs differ in who possesses voting authority, how extensively
governance rights are exercised, and how broadly voting power must be combined
to satisfy formal decision thresholds. The following sections examine whether
these differences are associated with token-market scale and token returns.

\section{Governance and Economic Outcomes}
\label{sec:governance-economic-outcomes}

Having established substantial variation in effective governance structure,
we next examine whether observable governance characteristics are associated
with economic outcomes. The analysis is intentionally observational. The
cross-sectional design can identify statistical associations between
governance structure and token-market scale, but it cannot establish whether governance
causes differences in valuation.

DAO governance tokens operate within programmable financial markets in which
control rights, protocol utility, and asset-market exposure may overlap
\cite{schar2021defi,Zetzsche2020}. Token-market scale is therefore treated as an observable size proxy rather than as a complete measure of protocol value or governance quality.

This distinction is particularly important because the direction of influence
may run both ways. Governance structure may affect the willingness of users,
delegates, or investors to commit capital to a protocol, but larger and more
mature DAOs may also develop broader delegate populations, different
participation patterns, or less concentrated voting-power distributions.
Unobserved differences in protocol age, sector, liquidity, token distribution,
or institutional maturity may additionally influence both governance structure
and market capitalization.

We therefore interpret the results below as cross-sectional associations rather
than estimates of a causal governance effect.

\subsection{Governance and Token-Market Scale}
\label{sec:governance-token-market-scale}

The primary token-market-scale analysis uses live-token FDV and contains 28 DAOs. The first
comparison asks whether scale is associated with governance measures that do
not mechanically divide by the number of delegates. Table~\ref{tab:governance-scale-rank}
reports the rank-based results.

\begin{table*}[t]
\centering
\caption{Spearman associations with live-token fully diluted valuation.}
\label{tab:governance-scale-rank}
\small
\begin{tabular}{lrrrr}
\hline
\textbf{Governance characteristic} & \textbf{$N$} & \textbf{$\rho$} &
\textbf{95\% CI} & \textbf{$p$} \\
\hline
$\ln(HHI)$ & 28 & $-0.124$ & $[-0.476,\;0.263]$ & 0.531 \\
Top-1 voting share & 28 & $-0.074$ & $[-0.436,\;0.308]$ & 0.707 \\
Top-10 voting share & 28 & $-0.329$ & $[-0.632,\;0.061]$ & 0.087 \\
$\ln(\text{Nakamoto coefficient})$ & 28 & 0.144 & $[-0.244,\;0.492]$ & 0.466 \\
$\ln(Q)$ & 20 & 0.031 & $[-0.417,\;0.467]$ & 0.897 \\
$\ln(\text{turnout})$ & 27 & 0.058 & $[-0.330,\;0.429]$ & 0.774 \\
$\ln(\text{delegates})$ & 28 & 0.545 & $[0.188,\;0.774]$ & 0.003 \\
$\ln(\text{token holders})$ & 28 & 0.612 & $[0.277,\;0.814]$ & $<0.001$ \\
\hline
\end{tabular}

\vspace{2pt}
\footnotesize
\parbox{\textwidth}{\textit{Note:} The outcome is live-token FDV based on mean September 2026
price and the governance-token supply retained by the matching pipeline after
automated verification or documented adjudication. $Q$ is the raw Effective
Quorum Coalition and is
defined only when current delegated voting power reaches a valid quorum.
Confidence intervals are bootstrap intervals reported by the analysis
pipeline.}
\end{table*}

No scale-free concentration measure has a statistically detectable association
with live-token FDV in the primary sample. The rank correlation for
$\ln(HHI)$ is $-0.124$ ($p=0.531$), and the Top-1 and Nakamoto-coefficient
results are similarly weak. The Top-10 share has a larger negative point
estimate, but its confidence interval includes zero. Raw quorum-coalition size
and supply-normalized turnout also show little association with scale.

By contrast, the breadth measures are strongly positive. The number of
registered delegates has $\rho=0.545$ ($p=0.003$), while the token-holder count
has $\rho=0.612$ ($p<0.001$). Larger token-market valuations therefore coincide
more consistently with larger governance and ownership populations than with
a more even distribution of voting power inside those populations.

\subsubsection{Regression and Bootstrap Evidence}

The bivariate HC3 regression of log live FDV on $\ln(HHI)$ produces

\begin{equation}
    \hat{\beta}_{HHI}=-0.461,
    \qquad SE=0.571,
    \qquad p=0.419.
\end{equation}

When delegate-base breadth is added, the coefficient on $\ln(HHI)$ is 0.823
($SE=0.828$, $p=0.320$), while the coefficient on
$\ln(Delegates)$ is 0.602 ($SE=0.224$, $p=0.007$). Adding turnout leaves the
substantive result unchanged: the HHI and turnout coefficients remain
statistically indistinguishable from zero, whereas the delegate-count
coefficient remains positive ($0.585$, $p=0.010$). Controls for token age and
Ethereum-mainnet deployment likewise do not yield an independent HHI effect.

The pairs bootstrap reinforces the uncertainty of the bivariate concentration
estimate. Across 1,000 resamples, the mean HHI coefficient is $-0.456$, with a
95\% percentile interval of $[-1.431,\;0.627]$. Although 81.4\% of slopes are
negative, the two-sided bootstrap $p$-value is 0.372. Directional consistency
across many resamples is therefore insufficient to establish a robust
concentration--scale relationship.

The modest sample also limits power. With $N=28$, the approximate minimum
detectable absolute rank correlation at 80\% power is 0.51. The absence of a
statistically detectable HHI association should not be read as proof that
moderate relationships are impossible.

\subsubsection{Alternative Token-Market-Scale Measures}

The alternative outcomes reveal why breadth and concentration must be modeled
jointly. With the CryptoCompare market-cap snapshot ($N=24$), the HHI rank
correlation is $-0.179$ ($p=0.402$), while delegate and holder counts remain
positive ($\rho=0.577$ and $0.642$, respectively). With trading-life-average
FDV ($N=44$), concentration appears more strongly in bivariate analysis:
$\rho=-0.292$ for $\ln(HHI)$ ($p=0.055$) and $\rho=-0.413$ for the Top-10 share
($p=0.005$). The corresponding bivariate OLS coefficient on $\ln(HHI)$ is
$-0.917$ ($SE=0.381$, $p=0.016$).

That apparent concentration effect disappears after accounting for the size
of the delegate base. In the trading-life-FDV model containing both variables,
the HHI coefficient becomes $-0.100$ ($p=0.808$), whereas the
$\ln(Delegates)$ coefficient is 0.452 ($p<0.001$). The same qualitative pattern
holds across the primary and market-cap outcomes: scale is associated with how
many participants the governance system contains, while conditional voting-
power dispersion is not robustly estimated.

\begin{figure*}[t]
    \centering
    \includegraphics[width=\textwidth]
    {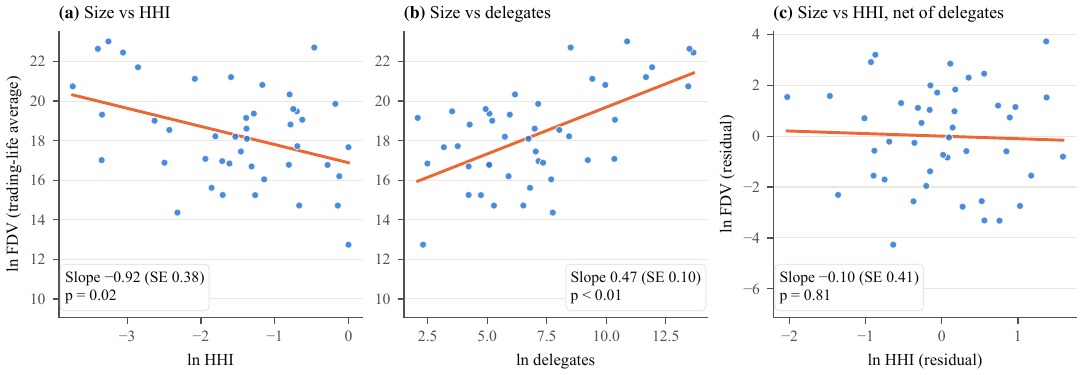}
    \caption{Token-market scale and voting-power concentration
    (trading-life-average FDV, $N=44$). Panel~(a) plots log FDV against log
    HHI; Panel~(b) against the log number of delegates; Panel~(c) shows the
    partial relation between log FDV and log HHI after both are residualized on
    log delegates. Lines are OLS fits; slopes, HC3 standard errors, and
    $p$-values are reported in each panel.}
    \label{fig:h1-scatter}
\end{figure*}

Figure~\ref{fig:h1-scatter} illustrates this pattern for the trading-life-FDV
sample.

The delegate participation rate and the delegate-normalized Quorum Coalition
Ratio are negatively correlated with several token-market-scale measures. Both variables
divide by the registered delegate count, which is itself positively associated
with scale. We therefore treat those correlations as partly mechanical and do
not interpret them as evidence that participation or quorum security declines
causally with token-market scale.

Taken together, the token-market-scale results support a bounded conclusion:

\begin{quote}
\textbf{Token-market scale is associated with the breadth of the observed
delegate and holder bases, but not with an independently robust reduction in
voting-power concentration.}
\end{quote}

This distinction motivates the return analysis. A governance system may grow
broader as a protocol matures without becoming proportionally less
concentrated, and neither property necessarily generates predictable token
returns.

\section{Governance and Token Returns}
\label{sec:governance-token-returns}

The preceding analysis indicates that token-market scale is more consistently
associated with the breadth of the delegate and holder bases than with the
concentration of delegated voting power itself. We next examine whether
governance characteristics are associated with token returns.

This question is distinct from whether governance is institutionally
important. Under the standard informational-efficiency benchmark, publicly
observable governance characteristics need not generate persistent abnormal
returns \cite{fama1970}. Cryptocurrency returns also exhibit market-specific
common factors and integrated trading relationships that must be separated
from governance-level variation \cite{liu2021,makarov2020}.

We consider three complementary forms of evidence: cross-sectional differences
in average returns, systematic exposure to Bitcoin, and short-run returns around
governance proposal completion. Across these analyses, the results provide
limited evidence that static governance characteristics independently explain
token-return variation.

\subsection{Cross-Sectional Average Returns}
\label{sec:cross-sectional-returns}

The return-history requirement yields 45 DAO tokens with at least 12 months of
observed monthly returns. Exact estimation samples are smaller where individual
governance variables are unavailable.

In the simplest raw-return specification, which contains 43 DAOs, the
coefficient on $\ln(HHI)$ is positive but statistically indistinguishable from
zero:

\begin{equation}
    \hat{\beta}_{HHI}
    =
    0.0142,
    \qquad
    p=0.214.
\end{equation}

Adding the delegate participation rate produces the same conclusion. In the
40-DAO complete-case sample, the coefficient on $\ln(HHI)$ is 0.0157
($p=0.216$), while the coefficient on log participation is $-0.0044$
($p=0.160$). Specifications that additionally incorporate quorum structure
likewise provide no statistically detectable relationship between the
governance variables and average monthly returns.

The same qualitative result holds when average returns are measured relative
to Bitcoin. Across the principal raw-return and BTC-relative specifications,
voting-power concentration, participation, and quorum-related measures do not
exhibit robust cross-sectional associations with average token performance.

A specification that additionally introduces end-of-sample live FDV produces
a positive and statistically significant coefficient on FDV, while the HHI
coefficient remains only marginally estimated. We do not treat this model as
primary evidence. Live FDV is measured near the end of a return history
extending back to 2021 and is therefore mechanically related to accumulated
past token-price performance. Conditioning historical average returns on this
later valuation can consequently induce a temporal relationship that should
not be interpreted as a governance or size effect.

The cross-sectional evidence therefore supports a relatively narrow
conclusion:

\begin{quote}
\textbf{Static governance characteristics do not robustly distinguish DAOs
with higher versus lower average token returns in the observed sample.}
\end{quote}

This result does not imply that governance is economically irrelevant.
Governance determines control rights, protocol modification, resource
allocation, and emergency authority. The evidence instead indicates that these
institutional differences do not translate into a simple cross-sectional
token-return relationship.

\subsection{Systematic Crypto-Market Exposure}
\label{sec:systematic-market-exposure}

DAO-token returns exhibit substantial contemporaneous exposure to the broader
crypto market. Across all 45 tokens with sufficient return histories,
comprising 1,853 DAO-month observations, the estimated Bitcoin coefficient is

\begin{equation}
    \hat{\beta}_{BTC}
    =
    0.907,
    \qquad
    SE=0.056,
\end{equation}

and the BTC-only model has $R^2=0.140$. Thus, DAO tokens in the sample move
strongly with Bitcoin on average, although Bitcoin alone explains only a
fraction of total monthly return variation.

\begin{figure}[t]
    \centering
    \includegraphics[width=\columnwidth]
    {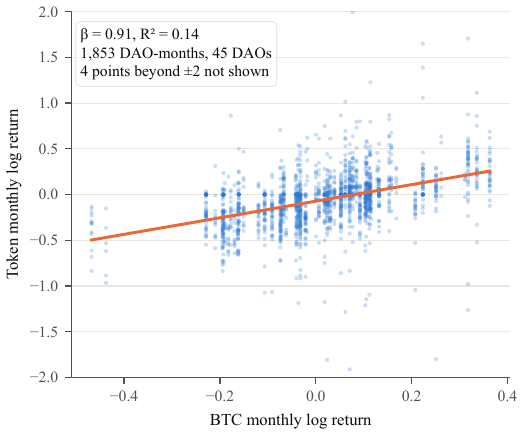}
    \caption{Monthly token log returns against Bitcoin log returns (45 DAOs,
    1,853 DAO-months). The line is the pooled OLS fit ($\beta=0.91$,
    $R^2=0.14$). Four observations with absolute monthly log return above 2 lie
    outside the plotted range.}
    \label{fig:btc-exposure}
\end{figure}

Figure~\ref{fig:btc-exposure} shows the pooled relation.

To evaluate the incremental contribution of governance variables, we next use
a common complete-case sample of 27 DAOs and 1,108 DAO-month observations. On
this common sample, the BTC-only specification has

\begin{equation}
    R^2=0.259.
\end{equation}

Adding $\ln(HHI)$ and $\ln(Delegates)$ increases explanatory power only
slightly,

\begin{equation}
    R^2=0.262,
    \qquad
    \Delta R^2 \approx 0.003.
\end{equation}

The HHI coefficient is not statistically distinguishable from zero under
either DAO-clustered inference ($p=0.270$) or the wild-cluster bootstrap
($p=0.296$). The delegate-count coefficient is more precisely estimated, but
its inference is borderline: $p=0.062$ under DAO clustering and $p=0.049$
under the wild-cluster bootstrap.

A further specification adding end-of-sample live FDV increases $R^2$ to
0.272 and produces a statistically significant HHI coefficient under both
clustered and wild-bootstrap inference. We do not interpret this as evidence
that governance concentration predicts historical monthly returns. As in the
cross-sectional analysis, live FDV is measured after much of the return history
used as the dependent variable and is mechanically related to prior token-price
performance. The scale-free specification therefore provides the more
defensible test of incremental governance information.

Taken together, the panel evidence indicates that contemporaneous crypto-market
movements account for substantially more return variation than the static
governance characteristics considered here. Governance variables add little
incremental explanatory power in the principal specification.

Importantly, this does not imply that Bitcoin explains most DAO-token return
variation. Even in the common sample, substantial residual variation remains.
The conclusion is comparative: within the variables examined here,
\emph{systematic market exposure is considerably more informative about
monthly token returns than static governance structure}.

\subsection{Governance Proposal Events}
\label{sec:governance-events}

The final proposal-event sample contains 1,806 concluded proposals from
35 DAOs with complete $[-5,+5]$ price windows. These proposals correspond to
1,391 distinct DAO--event dates and 19,866 event-window observations spanning
January 2021 through September 2026.

Across all proposal observations, the mean cumulative BTC-relative return over
the 11-day event window is

\begin{equation}
    \overline{CAR}[-5,+5]
    =
    -1.89\%.
\end{equation}

Approximately 60.7\% of proposal-level CARs are negative. Taken by itself,
this result suggests that DAO tokens tend to underperform Bitcoin around
proposal completion. However, raw CAR does not distinguish a proposal-specific
effect from persistent BTC-relative underperformance of the same DAO tokens.

We therefore compare each event CAR with the corresponding DAO's average
BTC-relative performance during complete non-event windows. On the full
proposal sample, the resulting mean excess CAR is

\begin{equation}
    \overline{ECAR}
    =
    -1.70\%.
\end{equation}

The point estimate remains negative, but its statistical interpretation depends
on how repeated and overlapping governance events are treated. With standard
errors clustered by DAO, the corresponding $p$-value is 0.065; two-way
clustering by DAO and calendar month gives $p=0.091$, while inference based on
one average observation per DAO gives $p=0.095$.

Table~\ref{tab:event-robustness} reports the principal dependence corrections.

\begin{table*}[t]
\centering
\caption{Governance proposal events and dependence-robust inference.}
\label{tab:event-robustness}
\small
\setlength{\tabcolsep}{6pt}
\begin{tabular}{lrrrr}
\hline
\textbf{Event sample} &
\textbf{Events} &
\textbf{Mean excess CAR} &
\textbf{DAO-cluster $p$} &
\textbf{DAO--month / DAO-level $p$} \\
\hline

All proposals
& 1,806
& $-1.701\%$
& 0.065
& 0.091 / 0.095 \\

Distinct DAO--event dates
& 1,391
& $-1.912\%$
& 0.043
& 0.063 / 0.102 \\

Non-overlapping event dates
& 736
& $-1.457\%$
& 0.137
& 0.174 / 0.192 \\

\hline
\end{tabular}

\vspace{2pt}
\footnotesize
\textit{Note:} Excess CAR is the proposal-window BTC-relative CAR minus the
corresponding DAO's average non-event-window CAR. The distinct-date
specification collapses same-DAO proposals ending on the same date.
The non-overlapping specification retains event dates separated sufficiently
to avoid overlapping $[-5,+5]$ windows within a DAO. DAO--month inference
clusters jointly by DAO and calendar month; DAO-level inference first averages
event effects within each DAO.
\end{table*}

The dependence corrections materially affect the inference. Collapsing
same-date events produces a somewhat larger negative point estimate and a
DAO-clustered $p$-value below 0.05, but the result is no longer conventionally
significant after DAO--month clustering or DAO-level aggregation. More
importantly, when overlapping within-DAO event windows are removed, the point
estimate remains negative but all three inference procedures yield
$p>0.10$.

A daily comparison leads to the same caution. Mean BTC-relative return on
event-window days is approximately $-0.240\%$, compared with $-0.226\%$ on
non-event days. The difference is economically small and statistically
indistinguishable from zero ($p=0.877$).

The event study therefore provides evidence of a negative \emph{point
estimate} around proposal completion, but not a robust governance-specific
return effect. The magnitude is approximately 1.5--2\% over the 11-day window
after accounting for each DAO's own non-event baseline, yet the statistical
evidence weakens once repeated events, common calendar shocks, and overlapping
windows are handled explicitly.

\begin{figure*}[t]
    \centering
    \includegraphics[width=\textwidth]
    {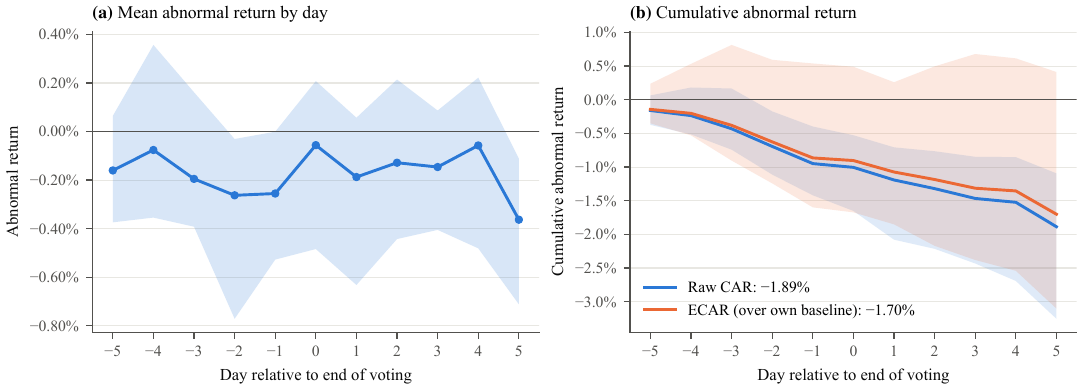}
    \caption{Abnormal returns around governance proposal events (1,806
    proposals, 35 DAOs). Panel~(a) reports the mean daily BTC-relative abnormal
    return from day $-5$ to day $+5$; Panel~(b) reports the cumulative raw CAR
    and the baseline-adjusted ECAR. Shaded bands are 95\% intervals from a
    bootstrap over DAOs (1,000 resamples).}
    \label{fig:event-path}
\end{figure*}

Figure~\ref{fig:event-path} shows the daily and cumulative abnormal-return
paths around proposal completion.

A 1,000-replication placebo analysis provides an additional diagnostic by
comparing the observed event statistic with event-free windows sampled from
the same DAOs. Because independently sampled placebo windows do not fully
preserve common calendar dependence, we treat the placebo distribution as a
complementary robustness check rather than as the principal inferential test.

\begin{figure*}[t]
    \centering
    \includegraphics[width=\linewidth]
    {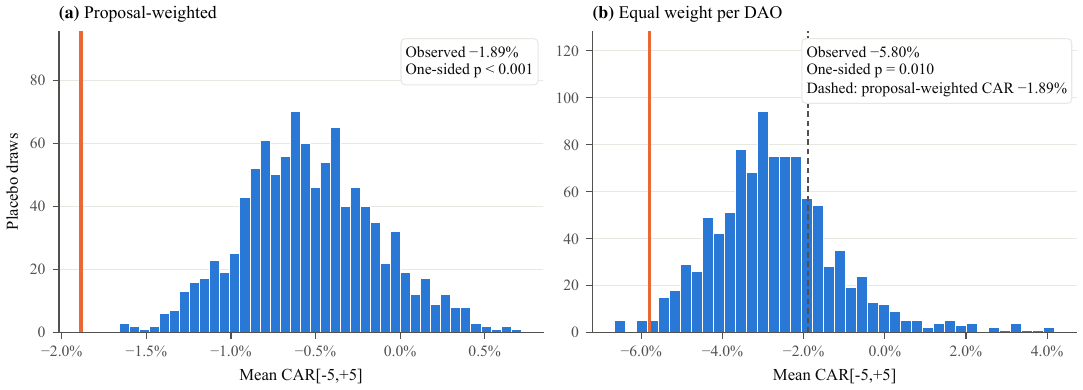}
    \caption{Raw-CAR placebo diagnostic for governance proposal events.
    The distribution is generated from 1,000 draws of event-free windows
    sampled within DAOs while preserving the DAO-level event-count structure;
    the observed raw BTC-relative CAR is shown for comparison. This diagnostic
    is not the principal inferential result because independently sampled
    windows do not preserve common calendar shocks or within-DAO overlap.
    Interpretation therefore relies on baseline-adjusted ECAR and the
    DAO-clustered, DAO--month-clustered, DAO-level, and non-overlapping
    specifications in Table~\ref{tab:event-robustness}.}
    \label{fig:event-car-placebo}
\end{figure*}

Overall, the event evidence does not support a simple interpretation in which
proposal completion systematically causes negative token-price reactions.
Negative point estimates are present across several specifications, but their
statistical significance is sensitive to event dependence and overlap.
Accordingly, the more defensible conclusion is that governance proposal
completion is associated with modest negative BTC-relative return estimates in
this sample, while evidence for a distinct and robust governance-event effect
remains weak.

\section{Governance Under Stress}
\label{sec:governance-under-stress}

The preceding sections characterize governance under routine operation:
authority is allocated through agenda access, voting and delegation, decision
thresholds, execution constraints, emergency authority, and external
dependencies, while the empirical analysis measures how voting authority is
distributed and exercised in practice.
Governance design becomes particularly consequential, however, when actors
behave strategically or when operational failures require action faster than
ordinary collective decision-making permits.

We distinguish two forms of governance stress. \emph{Adversarial governance}
concerns attempts to exploit voting power, delegation, agenda access, or
procedural rules to influence collective decisions. \emph{Emergency governance}
concerns failures for which the latency of ordinary governance may itself
become a source of risk. These problems are related but conceptually distinct: the former concerns misuse of governance authority, whereas the latter concerns whether legitimate authority can respond sufficiently quickly to protect the system. Accordingly, resilience in this section refers to the design capacity of governance arrangements to absorb, escalate, pause, replace, or reverse decisions under adversarial or emergency conditions. It is analyzed through mechanism design and illustrative incidents, rather than through population-level estimates of failure frequency or response latency.

The governance metrics introduced earlier should therefore be interpreted as
measures of \emph{structural exposure}, not as evidence that an attack or
failure has occurred. Documented incidents provide a separate evidence layer
for examining how governance mechanisms behave when these exposures are
realized.

\subsection{Structural Exposure to Governance Capture}
\label{sec:governance-capture}

Governance capture can occur whenever a comparatively narrow set of actors
obtains sufficient influence over one or more stages of the governance
lifecycle to steer outcomes, block corrective action, or determine which
issues reach binding decision-making
\cite{feichtinger2024sokdaos,fritsch2024votingpower,weidener2025delegated,
gudgeon2020crisis}.
Capture need not depend solely on direct token ownership. Effective influence
may arise through delegation, agenda-setting authority, coordinated voting,
temporary economic incentives, or privileged execution roles.

The empirical concentration measures provide one indication of this exposure.
A high HHI or top-$k$ share means that a comparatively small number of delegates
control a large fraction of delegated voting power. Similarly, a small
Effective Quorum Coalition indicates that relatively few of the largest
delegates are arithmetically sufficient to satisfy the formal quorum
requirement. In the present sample, the prevalence of single-delegate quorum
sufficiency illustrates why formal participation thresholds cannot be
interpreted independently of the distribution of delegated voting power.

These measures do not establish capture. A large delegate may vote
independently, and several large delegates need not coordinate. Likewise,
satisfying quorum is not equivalent to satisfying all proposal-passage
requirements. Structural concentration therefore identifies where coordinated
control may require fewer observed delegate positions, but
behavioral or event-level evidence is required to establish an actual
governance attack.

A second exposure arises through temporary coordination. Bribery, vote buying,
or other proposal-specific incentives can concentrate effective voting
behavior without producing persistent ownership or delegation concentration
\cite{feichtinger2024sokdaos}. Such behavior cannot be inferred reliably from
HHI alone and instead requires proposal-level evidence such as documented
incentive arrangements, unusual delegation changes, or explicitly coordinated
voting behavior.

Governance capture can also occur before formal voting begins. Proposal
thresholds, sponsorship requirements, facilitator processes, and informal
forum norms determine which issues can progress to a binding vote. Consequently,
agenda access may remain concentrated even when voting eligibility is formally
broad. This creates a distinct governance surface in which control is exercised
through proposal admission rather than through the final vote itself.

Taken together, capture exposure is therefore distributed across the
governance lifecycle:

\begin{equation}
\text{Agenda}
\rightarrow
\text{Voting}
\rightarrow
\text{Thresholds}
\rightarrow
\text{Execution}
\rightarrow
\text{Exceptional control}.
\end{equation}

Here, exceptional control includes emergency authority and governance responses
triggered by external-dependency failures, such as oracle, bridge, or feed
disruption. A system can be comparatively decentralized at one stage and
concentrated at another. Governance resilience should therefore be evaluated
across these stages rather than inferred from voting-power distribution alone.

The reviewed literature documents bribery, temporary voting-power acquisition,
and procedural manipulation as feasible governance attack paths
\cite{feichtinger2024sokdaos}. We do not infer their realization from
concentration alone. Classifying a proposal as manipulated would require
proposal-specific evidence---for example, a documented side payment, a
temporary delegation or borrowing trace linked to the vote, or a verified
procedural exploit. Such behavioral labels are outside the present
cross-sectional dataset, which measures structural exposure only.

\subsection{Emergency Governance as a Response Layer}
\label{sec:emergency-governance}

Emergency governance addresses the opposite problem: legitimate collective
authority may be too slow to contain a rapidly developing operational or
security failure. Ordinary governance typically requires proposal formation,
voting, quorum satisfaction, and possibly delayed execution. During a
time-sensitive incident, the latency of this process may exceed the time
available to prevent additional damage.

The design-space analysis identified several mechanisms intended to address
this problem, including pause and freeze controls, circuit breakers, emergency
multisignatures, security councils, and shutdown or settlement procedures.
Rather than repeating their technical definitions here, we evaluate them as
alternative mechanisms for reallocating authority under stress.

The central governance trade-off is that reducing response latency generally
requires transferring some authority from the broad governance population to
a smaller actor, committee, or automated rule. The relevant question is
therefore not simply whether emergency authority exists, but how constrained
that authority is.

Five properties are particularly important:

\begin{itemize}
    \item \textbf{Scope:} which operations or protocol components can be
    affected;

    \item \textbf{Activation:} whether intervention is discretionary,
    threshold-based, or automatic;

    \item \textbf{Authority structure:} whether control is held by a single
    actor, multisignature, council, or predefined mechanism;

    \item \textbf{Reversibility and duration:} whether the intervention is
    temporary, revocable, or effectively terminal; and

    \item \textbf{Accountability:} whether the action is auditable and subject
    to subsequent token-holder review, ratification, replacement, or reversal.
\end{itemize}

These properties determine whether emergency authority functions as a narrowly
scoped safety layer or becomes a persistent alternative center of governance
power. Deliberate crisis-resolution mechanisms can therefore improve
responsiveness, but they also introduce privileged-authority and accountability
trade-offs \cite{Shevko2023}. When emergency authority is exercised through
committees, security councils, or multisignature bodies, the rule by which
members are selected becomes part of the governance design: deterministic
fairness and bounded-influence results for committee selection provide a formal
benchmark for assessing how much adversarial weight would be required to capture
the decision body \cite{Melnikov2026}. The desirable configuration is therefore
not necessarily the absence of privileged authority, but an architecture in
which exceptional authority is limited to the actions and duration necessary to
contain the corresponding risk.

This relationship can be expressed as

\begin{quote}
\textbf{Operational resilience may require concentrated emergency authority,
but the governance risk created by that concentration depends on scope,
duration, auditability, and revocability.}
\end{quote}

Execution constraints and emergency authority are closely connected. A
timelock provides time during which an approved but unsafe action can be
detected; an emergency mechanism determines whether anyone possesses the
authority to act on that information before execution. Conversely, an
emergency mechanism with broad discretionary powers but weak procedural
constraints may itself constitute a governance risk.

Accordingly, the comparative coding treats emergency authority as a vector of
properties rather than a binary indicator: actor, permitted action, activation
condition, signer threshold, duration, revocation, auditability, and
post-incident review. This avoids treating a narrowly scoped pause role as
equivalent to an unrestricted upgrade key and makes version-specific changes
to emergency powers visible.

\subsection{Illustrative Stress Cases}
\label{sec:incident-case-studies}

Documented incidents provide evidence about the behavior of governance
mechanisms under conditions that cannot be inferred from architecture alone.
At the present stage, however, the incident evidence is not a systematically
enumerated population of DAO failures. We therefore treat the following cases
as \emph{illustrative stress cases} rather than as observations from which the
frequency or relative effectiveness of governance mechanisms can be estimated.

\paragraph{MakerDAO --- Black Thursday (March 2020)}
The market disruption of 12--13 March 2020 combined extreme collateral-price
movements with failures in the liquidation process, including zero-bid
auctions. The incident demonstrates that governance resilience depends not
only on formal voting architecture but also on operational parameters,
liquidation mechanisms, external market conditions, and the ability of
governance to recapitalize the system and adjust risk controls after failure
\cite{makerwhitepaper2020,schar2021defi}.
The protocol's incident review documents the interaction among market stress,
network congestion, keeper participation, and auction behavior.
\footnote{MakerDAO, ``The Market Collapse of March 12--13, 2020: How It
Impacted MakerDAO,'' available at
\url{https://blog.makerdao.com/}.}
We use the event to identify interacting failure surfaces, not to attribute
every later governance change to a single incident.

\paragraph{Compound --- COMP distribution bug (September 2021)}
A governance-approved upgrade introduced an error in the COMP distribution
mechanism that resulted in unintended token distributions. The incident
illustrates an important distinction between emergency intervention and state
reversibility: privileged pause or administrative authority does not
necessarily permit already executed state transitions to be undone. Compound's
contemporaneous governance digest links the fault to Proposal~62 and records
the subsequent patch process.
\footnote{Compound, ``Compound Treasury Updates, COMP Bug Fix, Dynamic Risk
Parameters,'' 14 October 2021, available at
\url{https://compound.substack.com/p/compound-treasury-updates-comp-bug}.}
The case therefore illustrates both code-review risk and the latency imposed by
the same governance process needed to authorize remediation.

\paragraph{Aave --- v2/v3 security incident (November 2023)}
Aave temporarily paused affected markets following identification of a
vulnerability, after which governance processes were used to manage recovery
and restoration \cite{aave2023incident}. This case illustrates the intended
division of labor between emergency and ordinary governance: narrowly scoped
authority can provide rapid containment while broader governance retains
responsibility for subsequent remediation and normalization.

The three cases demonstrate different failure modes. MakerDAO illustrates a
stress event involving market, oracle, liquidation, and risk-management
dependencies; Compound illustrates the difficulty of reversing a governance
action after execution; and Aave illustrates the use of delegated emergency
authority for rapid containment. Their value is comparative rather than
statistical.
They were selected because they expose distinct boundary conditions---market
and auction stress, faulty governance execution, and rapid containment---not
because they constitute a complete incident population. We therefore make no
claim about the relative frequency of these failure modes or the average
effectiveness of the corresponding controls.

\subsection{External Dependencies and Oracle Governance}
\label{sec:oracle-dependencies}

External dependencies can transform a technical failure into a governance
problem when recovery requires discretionary or time-sensitive intervention.
Oracle systems provide an important example because smart contracts cannot
directly observe external state and therefore depend on mechanisms that
introduce prices or other off-chain information
\cite{eskandari2021sokOracles}.

From a governance perspective, the relevant questions concern authority:
who selects the feed, who can modify its parameters, what fallback source is
available, who can pause or replace the oracle, and how disputes or failures
are escalated. When oracle signers, feed providers, or replacement committees
are selected through governance-controlled procedures, the same bounded-influence
logic applies to the selection of data-authority roles as to emergency
committees. These choices define a control boundary between on-chain
governance and the external information on which protocol execution depends.

A stale, manipulated, or unavailable oracle can affect liquidations,
collateral valuation, issuance, or other safety-critical state transitions
before ordinary governance can react. Resilience therefore depends on both
the design of the external data mechanism and the governance response available
when that mechanism fails. Responses may include switching to a fallback feed,
constraining affected operations, activating a circuit breaker, or pausing a
market until reliable data are restored.

Oracle governance thus illustrates a broader principle:

\begin{quote}
\textbf{External dependencies become governance dependencies when their
selection, failure, or recovery changes who must exercise authority over
protocol operation.}
\end{quote}

For this reason, the SoK treats oracle systems as part of the broader governance
and security boundary rather than as an independent governance domain.
The comparative coding is correspondingly limited to governance-relevant
properties: feed-selection authority, fallback mechanisms, pause or
replacement authority, dispute pathways, and escalation to token-holder or
emergency governance. It does not attempt to benchmark oracle accuracy or
latency independently of the consuming protocol.

\section{Synthesis}
\label{sec:synthesis}

The preceding analyses examine DAO governance at three complementary levels.
The governance-design analysis identifies where authority is formally located
across six recurring layers: agenda access, voting and delegation, decision
thresholds, execution constraints, emergency authority, and external
dependencies. The empirical analysis measures how voting authority is
distributed and exercised in deployed governance systems. The stress analysis
then considers how these institutional arrangements behave when governance is
challenged by strategic behavior, operational failure, or time-sensitive
dependency breakdowns.

Taken together, the evidence suggests that DAO decentralization is not a
binary property and cannot be inferred from the existence of token voting
alone. Governance is better understood as a layered allocation of authority
whose degree of decentralization depends on both formal institutional design
and the realized distribution of control.

\subsection{From Formal Governance to Effective Governance}
\label{sec:formal-effective-governance}

The central distinction emerging from the combined analysis is between
\emph{formal governance} and \emph{effective governance}.

Formal governance specifies which actors are institutionally permitted to
propose, vote, delegate, execute, or intervene. These rights and constraints
are encoded through proposal thresholds, governor contracts, quorum and
passage rules, timelocks, execution paths, and emergency authority.

Effective governance concerns how authority is distributed and exercised
within those formal rules. A system may permit thousands of addresses to vote
or delegate while a comparatively small number of delegates control most
voting power. Likewise, a formally demanding quorum may nevertheless be
satisfied by one or a few highly weighted actors, while broad formal voting
eligibility may coexist with limited realized participation.

The principal implication is therefore:

\begin{quote}
\textbf{Formal decentralization does not necessarily imply a distributed
allocation of effective decision-making authority.}
\end{quote}

The empirical evidence provides a concrete illustration of this distinction.
Across the observed Tally-indexed DAOs, the median largest delegate controls
approximately 42\% of delegated voting power, and the ten largest delegates
collectively control approximately 98\%. Among the 38 DAOs with a valid EQC,
the median coalition size is one, and approximately 63\% exhibit SDQS.

These observations indicate arithmetic quorum sufficiency, not coordination,
capture, or unilateral proposal passage. They instead show that broad formal
participation rights can coexist with a highly concentrated distribution of
structural voting capacity.

Effective decentralization should therefore be evaluated through several
distinct questions:

\begin{enumerate}
    \item \emph{Who can initiate governance action?}
    \item \emph{Who possesses effective voting authority?}
    \item \emph{Who actually participates?}
    \item \emph{How broadly must authority be combined to satisfy formal
    decision thresholds?}
    \item \emph{Who controls execution and exceptional intervention, and which
    external dependencies can trigger such intervention?}
\end{enumerate}

These questions correspond to different governance layers and should not be
collapsed into a single decentralization score. A DAO may be comparatively
open at one layer and highly concentrated at another.

\subsection{Recurring Governance Trade-offs}
\label{sec:governance-tradeoffs}

The systematization reveals several recurring trade-offs that arise because
governance mechanisms frequently solve one coordination problem while creating
another.

\paragraph{Openness versus gatekeeping}
Lower proposal barriers broaden agenda access but can increase spam,
procedural congestion, and strategic use of governance. Higher thresholds,
sponsorship requirements, or mediated proposal processes filter these costs
but can concentrate agenda-setting authority among large token holders,
prominent delegates, or governance intermediaries.

\paragraph{Representation versus concentration}
Delegation reduces the participation burden on individual token holders and
permits specialized representatives to monitor governance continuously.
However, delegated voting power can accumulate around a comparatively small
number of prominent actors
\cite{fritsch2024votingpower,weidener2025delegated}. Delegation can therefore
increase representative capacity while simultaneously increasing effective
voting-power concentration.

\paragraph{Decision accessibility versus coalition breadth}
Lower quorum and passage requirements reduce the risk of governance
stagnation, but concentrated voting power may allow formal participation
requirements to be satisfied by a narrow coalition. Higher thresholds can
require broader participation while increasing the probability of failed or
blocked governance actions.

\paragraph{Responsiveness versus execution safety}
Timelocks and review windows provide time to inspect approved actions before
execution, but they also delay legitimate state changes. During security
incidents, the same delay that protects against unsafe execution may impede
rapid remediation.

\paragraph{Decentralized legitimacy versus emergency responsiveness}
Guardians, security councils, multisignatures, and automated emergency authorities
can react faster than ordinary token-holder governance. Their operational
advantage is obtained by concentrating exceptional authority in a smaller set
of actors. The governance consequences therefore depend on whether this
authority is narrowly scoped, auditable, revocable, time-limited, and subject
to subsequent review.

\paragraph{Functionality versus dependency exposure}
Oracles, bridges, cross-chain messengers, and other external data or execution
dependencies extend protocol functionality, but they also introduce failure and
control points that may require governance intervention. More redundant or
decentralized dependency designs can reduce single points of failure, but they
may increase complexity, cost, monitoring burden, and coordination overhead.
The relevant governance question is therefore not whether an external dependency
exists, but who selects it, how failures are detected, how fallbacks are
activated, and how authority is escalated when reliable operation is lost.

These trade-offs imply that no governance mechanism should be classified as
intrinsically ``decentralizing'' or ``centralizing'' in isolation. Its effect
depends on how it redistributes authority across the broader governance
architecture.

\subsection{Governance Design--Measurement--Outcome Matrix}
\label{sec:governance-matrix}

The principal SoK contribution is to connect formal governance mechanisms with
observable measures of effective authority, plausible failure exposures,
empirical or operational evidence, and corresponding design responses.
Table~\ref{tab:integrated-framework} summarizes this mapping.

The table distinguishes \emph{structural exposure} from \emph{realized
failure}. For example, a high HHI or a small Effective Quorum Coalition
indicates concentrated governance capacity but does not establish that
delegates coordinate or that a governance attack has occurred. Similarly,
emergency authority creates privileged control but does not imply misuse.

\begin{table*}[t]
\centering
\caption{Integrated governance design, measurement, failure-exposure, and
outcome framework.}
\label{tab:integrated-framework}
\scriptsize
\setlength{\tabcolsep}{2pt}
\renewcommand{\arraystretch}{1.18}

\begin{tabular}{
p{0.13\textwidth}
p{0.14\textwidth}
p{0.17\textwidth}
p{0.15\textwidth}
p{0.22\textwidth}
p{0.14\textwidth}}
\hline

\textbf{Dimension} &
\textbf{Mechanism} &
\textbf{Observable Measure / Evidence} &
\textbf{Failure Exposure} &
\textbf{Evidence in This Study} &
\textbf{Design Response} \\
\hline

Agenda access
&
Proposal threshold, sponsorship, staged or mediated proposal process
&
Threshold level; proposer eligibility; sponsorship and process requirements
&
Agenda capture; spam; procedural gatekeeping
&
Systematized qualitatively; not directly measured in the cross-DAO empirical
analysis
&
Transparent admission rules; calibrated thresholds; alternative sponsorship
paths
\\[3pt]

Voting distribution
&
Token-weighted voting and delegation
&
HHI; Top-1/5/10 shares; governance Nakamoto coefficient; delegate coverage
&
Voting-power concentration; structural capture exposure
&
High concentration is common; median Top-1 share is approximately 42\% and
median Top-10 share approximately 98\%. Concentration is not robustly
associated with primary token-market scale after accounting for delegate-base
breadth
&
Concentration monitoring; transparent delegation; delegate accountability
\\[3pt]

Participation
&
Direct and delegated voting
&
Delegate participation rate; supply-normalized turnout
&
Inactive electorate; reliance on a narrow set of active participants
&
Participation varies substantially across DAOs. Static participation measures
do not robustly explain average token returns
&
Participation monitoring; delegate accountability; incentive mechanisms where
empirically justified
\\[3pt]

Quorum
&
Quorum and passage thresholds
&
Effective Quorum Coalition $Q_i$; Quorum Coalition Ratio
&
Narrow coalition sufficient to satisfy formal participation requirements;
quorum blocking
&
Among DAOs with a valid EQC, median $Q_i=1$ and approximately 63\%
exhibit SDQS
&
Calibrated thresholds; monitoring of coalition structure; proposal-specific
rules where appropriate
\\[3pt]

Execution
&
Timelock, queue, staged execution
&
Execution delay; review window; cancellation or veto path; executor structure
&
Insufficient review at low delay; slow remediation at high delay
&
Governance-design and incident evidence identify a
responsiveness--execution-safety trade-off; no cross-DAO economic coefficient
is estimated
&
Risk-calibrated delay; observable queued actions; cancellation or intervention
path
\\[3pt]

Emergency authority
&
Guardian, security council, emergency multisignature
&
Actor composition; permitted actions; signer threshold; duration; revocation
&
Privileged-power misuse; committee or key concentration
&
Illustrative incidents show both rapid containment and limits of privileged
intervention; the current incident set is not statistically representative
&
Narrow scope; auditability; time limits; revocation; post-incident governance
review
\\[3pt]

External dependencies
&
Oracle feeds, fallback mechanisms, dispute systems
&
Feed-selection authority; fallback path; replacement or pause authority;
dispute mechanism
&
Manipulated or stale inputs; liveness failure; dispute capture
&
Treated as a governance and security dependency whose failure may require
emergency intervention; no independent cross-DAO economic effect is estimated
&
Fallback feeds; circuit breakers; transparent escalation and dispute
resolution
\\

\hline
\end{tabular}

\vspace{3pt}
\footnotesize
\textit{Note:} Failure exposure denotes a structural vulnerability or potential
failure mode and does not imply that a corresponding attack or governance
failure occurred. Empirical associations are observational and should not be
interpreted causally.
\end{table*}

\subsection{A Layered View of DAO Decentralization}

The integrated framework shifts the analytical question from
``Is this DAO decentralized?'' toward a more precise sequence of questions:

\begin{quote}
\emph{Where is authority located? How concentrated is it in practice?
Who exercises it? How broadly must it be combined to make decisions?
Who controls execution and exceptional intervention, and which external
dependencies can trigger such intervention?}
\end{quote}

This layered perspective reconciles several observations that otherwise appear
contradictory. A DAO can offer broad formal voting access while exhibiting
concentrated delegated power. Delegation can improve representation while
increasing concentration. A demanding quorum can coexist with
single-delegate quorum sufficiency. Timelocks can increase execution safety while reducing responsiveness. Emergency councils can improve resilience while introducing privileged authority, especially when committee-selection rules do not bound adversarial influence.

DAO decentralization is therefore better understood as a configuration of
authority across governance layers than as a single scalar property. The main
analytical task is not to assign a universal decentralization score, but to
identify where authority is concentrated, how that concentration interacts
with formal governance rules, and which safeguards constrain its exercise.

\section{Discussion}
\label{sec:discussion}

DAO governance generates unusually rich public audit trails: proposal rules,
delegation relationships, voting weights, quorum parameters, execution paths,
and many privileged roles can often be inspected through contracts, governance
platforms, and public records. That
observability makes authority measurable; it does not make authority broadly
distributed. The findings therefore reinforce the paper's central distinction:
governance design, effective authority, token-market outcomes, and operational
resilience are related but analytically distinct properties.

\subsection{What the Empirical Evidence Adds to the SoK}
\label{sec:discussion-decentralization}

The systematized design space identifies where authority \emph{can} reside;
the empirical analysis shows where voting authority \emph{does} reside in the
observed systems. Large formal delegate populations can coexist with highly
concentrated voting power, and an apparently demanding quorum can be
arithmetically reachable by one large delegate position.

The measures must nevertheless remain distinct. HHI and top-$k$ shares describe
the distribution of voting capacity, participation measures describe its use,
and the Effective Quorum Coalition links the observed distribution to a formal
threshold. A DAO can score differently on each dimension, so no single measure
is a complete decentralization index.

\subsection{Governance and Token-Market Outcomes}
\label{sec:discussion-governance-returns}

The economic analysis separates the institutional importance of governance
from its usefulness as a token-market signal. Larger token-market valuations
coincide with broader delegate and holder bases, while voting-power
concentration has no robust independent association after delegate breadth is
included. This relationship is associational: maturity, liquidity, adoption,
sector, and token distribution may influence both governance breadth and
valuation.

Static governance characteristics likewise do not robustly distinguish
higher-return from lower-return DAO tokens. Contemporaneous Bitcoin returns
explain substantially more monthly variation, and negative proposal-event
point estimates weaken after repeated events, overlap, and common calendar
dependence are addressed.

The resulting interpretation is:

\begin{quote}
\textbf{Governance can be institutionally and operationally consequential
without generating a simple or robustly predictable short-run token-return
premium.}
\end{quote}

Governance determines control rights, protocol modification, treasury
authority, delegation, execution, and incident response. These functions remain
important even when their effects do not appear as stable return predictability.

\subsection{Transparency, Observability, and Accountability}
\label{sec:discussion-transparency}

Blockchain-based governance provides substantial transparency because proposal
histories, voting records, delegation relationships, contract parameters, and
many execution actions are publicly inspectable
\cite{tallydocs,snapshotdocs}. This makes structural governance properties
measurable in ways that are often difficult in conventional organizations.

Transparency, however, should be distinguished from decentralization. Public
delegation records can reveal concentrated voting power without reducing that
concentration. Public voting histories can reveal inactivity without increasing
participation. Public emergency transactions can make privileged intervention
auditable without eliminating the privilege itself.

A more useful interpretation is therefore:

\begin{quote}
\textbf{Transparency determines how observable governance authority is;
decentralization concerns how that authority is distributed and constrained.}
\end{quote}

This distinction also affects the interpretation of governance risk.
Concentration measures and public governance histories can identify structural
exposure, but they do not demonstrate collusion, bribery, capture, or malicious
coordination. Transparency improves the ability to monitor such risks; it does
not by itself remove them.

\subsection{Implications for DAO Design}
\label{sec:discussion-designers}

The combined evidence suggests that governance monitoring should focus on
effective authority rather than nominal participation capacity. Delegate
counts, token-holder counts, and open voting eligibility are useful
descriptions of governance breadth, but they should be complemented by
voting-power concentration, top-$k$ shares, participation measures, and quorum
coalition structure.

These measures are most useful as a dashboard rather than as a composite
score. A single decentralization index would conceal important differences
between concentration, inactivity, agenda control, execution authority, and
emergency intervention.

Delegation deserves particular attention because it can improve representative
capacity while simultaneously concentrating effective authority. Governance
systems using delegation should therefore make large delegation positions,
voting histories, voting rationales, conflicts of interest, and re-delegation
options readily observable. The objective is not to eliminate professional
delegation, but to make concentrated representative authority accountable and
contestable.

Execution and emergency authority should likewise be treated explicitly.
Timelocks, guardians, security councils, multisignatures, and pause mechanisms
solve different problems and should have clearly defined boundaries. Where
privileged intervention is necessary, narrow scope, auditability, revocability,
time limitations, and subsequent governance review reduce the risk that an
emergency mechanism becomes an unconstrained parallel governance system.

Finally, governance metrics should be monitored longitudinally. Delegated
voting power, active participation, and quorum sufficiency can change as
delegations move and governance populations evolve. A system that appears
dispersed at one observation date may become substantially more concentrated
without any formal change to the governance contract.

\subsection{Implications for Research}
\label{sec:discussion-research}

The empirical limitations of the present study identify several priorities for
future DAO-governance research.

First, governance measurement should become longitudinal. Historical delegate
distributions would permit HHI, top-$k$ shares, participation, and quorum
coalition structure to be reconstructed at the time economic or governance
events occur rather than applying a later governance snapshot retrospectively.

Second, participation should be studied at multiple levels. Address-based
participation, voting-power turnout, delegate entry and exit, re-delegation,
and proposal salience capture different dimensions of realized governance
activity.

Third, proposal-event studies would benefit from reproducible proposal
taxonomies. Treasury allocations, parameter changes, protocol upgrades,
governance-process changes, and emergency proposals may have very different
information content, and pooling them can obscure heterogeneous effects.

Fourth, comparative research on governance resilience requires a systematic
incident database. Governance incidents are currently dispersed across
post-mortems, governance forums, security disclosures, transaction histories,
and proposal archives. A standardized dataset recording the affected
governance layer, detection mechanism, emergency actor, authority exercised,
response latency, reversibility, follow-up, and documented consequence would
make it possible to move from illustrative cases toward comparative evidence.

Fifth, formal analysis should complement empirical measurement. Deterministic
committee-selection bounds, mechanism-design models of delegation and bribery,
and verifiable emergency-authority policies can translate structural exposure
into testable safety conditions \cite{Melnikov2026}. Such formal results do not
replace on-chain measurement, but they provide a useful benchmark for assessing
whether a governance design can guarantee honest-majority or bounded-influence
properties under stated adversarial assumptions.

More generally, DAO-governance research should distinguish explicitly among
formal rules, structural exposure, observed behavior, and realized outcomes.
Conflating these levels can transform a concentration metric into an implied
attack measure, a governance snapshot into a historical causal variable, or
token returns into an overly broad test of governance quality.

\section{Limitations}
\label{sec:limitations}

The combined SoK and empirical design broadens the analysis of DAO governance,
but several limitations constrain the scope and interpretation of the results.
These limitations concern literature coverage, platform and market-data
selection, measurement, temporal alignment, and statistical identification.

\subsection{Literature and Protocol Coverage}
\label{sec:limitations-literature}

DAO governance evolves rapidly, and relevant evidence is distributed across
peer-reviewed research, protocol documentation, governance forums, proposal
archives, technical repositories, and incident reports. The structured SoK
strategy is designed to saturate the design space, not to provide exhaustive
bibliometric coverage; relevant mechanisms or protocol versions may therefore
remain outside the corpus.

Protocol-specific evidence also varies in quality and preservation. Governance
systems with detailed documentation, active forums, and archived proposals are
easier to reconstruct than systems with fragmented or changing records.
The resulting design space should therefore be interpreted as a structured
representation of the observable evidence corpus rather than an exhaustive
enumeration of every DAO governance mechanism. The purposive protocol cases
also favor mature systems with unusually rich public records, so their
documentation quality should not be treated as typical of the DAO population.

\subsection{Platform and Sample Selection}
\label{sec:limitations-platform}

The empirical universe is derived from governance organizations indexed by
Tally and subsequently filtered using governance-token and CryptoCompare
market-data availability. The sample therefore favors EVM-compatible,
Governor-style on-chain governance for which standardized delegate, proposal,
and quorum information can be retrieved programmatically.

Governance conducted through other ecosystems, custom contracts, multisignature
coordination, or predominantly off-chain platforms such as Snapshot may be
underrepresented or absent. In addition, some protocols operate several
governance systems, while the governance instance represented in Tally may not
capture the complete institutional structure of the protocol.

The empirical findings should therefore be interpreted as evidence from the
observed Tally-indexed population rather than as estimates for the complete
universe of DAOs.

\subsection{Token Matching and Market-Data Availability}
\label{sec:limitations-token-matching}

Matching governance tokens to CryptoCompare introduces an additional selection
and measurement boundary. Ticker symbols are not unique, and the market-data
interface used in the current pipeline does not provide reliable contract
addresses for every asset. Candidate matches are therefore validated using
project identity and, where available, comparisons between governance-token
supply and CryptoCompare supply.

Forty-two of the 54 final token matches satisfy the automated supply-consistency
criterion, while 12 rely on documented manual adjudication because a comparable
supply value is unavailable or unsuitable for automatic verification. These
cases should be treated as auditable matching decisions rather than as
algorithmically verified observations. Each adjudication and its rationale is
preserved in the replication package, but the decisions have not been validated
by an independent market-data provider.

Market-data availability also varies substantially. Of the 54 DAOs in the
governance sample, only 46 have usable daily price data and 45 have at least
12 months of return history. The primary live-FDV analysis is smaller still.
Consequently, the economic analyses describe subsets of the governance sample,
and sample composition can influence the estimated relationships.

\subsection{Survivorship and Observability}
\label{sec:limitations-survivorship}

The design conditions on DAOs that remain observable through the selected
governance and market-data sources. Failed, abandoned, delisted, inactive, or
migrated organizations may be less likely to appear in the final economic
sample.

If governance structure is related to organizational survival or continued
market observability, this creates survivorship bias. The direction of that
bias cannot be inferred from the observed sample alone. The reported
governance--outcome relationships should therefore be interpreted conditional
on inclusion in the observable Tally--CryptoCompare population.

\subsection{Static Governance Measurement}
\label{sec:limitations-static}

The principal governance variables are measured primarily from the governance
state observed in October 2026, whereas token-return and proposal histories
extend from January 2021 through September 2026. Voting-power concentration,
delegate counts, and related measures are therefore later cross-sectional
snapshots rather than contemporaneous historical governance states.

This temporal mismatch is especially important for return analysis. Repeating
an October 2026 governance observation across historical DAO-months does not
imply that the same governance structure existed throughout the return period.
The panel consequently provides evidence about systematic market co-movement
and cross-sectional differences associated with the observed governance state,
not about the effect of contemporaneous governance changes on returns.

The same limitation affects causal interpretation of token-market scale.
Governance may affect organizational development, organizational development
may alter governance, or both may respond to common factors. Longitudinal
governance snapshots would be required to establish temporal ordering more
credibly.

\subsection{Delegate Distribution and Coverage}
\label{sec:limitations-hhi}

Delegate retrieval includes up to 500 addresses, while the primary HHI is
reported for the 100 largest delegates. Top-100 coverage is high in the
observed sample: the mean is approximately 98.9\%, the median is approximately
100\%, and the lowest observed value is approximately 87.3\%. HHI calculated
from all collected delegates is therefore available as a direct robustness
comparison.

Nevertheless, the delegate distribution remains dependent on the consistency
of Tally's aggregate delegated-vote total and returned delegate records. A
small number of coverage values marginally exceed one, indicating minor
inconsistencies between the aggregate denominator and the sum of individual
delegate records. These deviations are small but illustrate that API-derived
governance measures are not measurement-error free.

The concentration results should therefore be interpreted comparatively and
checked against HHI based on all collected delegates and complementary top-$k$
measures rather than relying on a single concentration statistic.

\subsection{Participation Measurement}
\label{sec:limitations-participation}

The delegate participation rate combines a historical numerator with a current
denominator. The numerator is the mean number of unique voters across recent
concluded proposals, whereas the denominator is the current registered-delegate
count. Delegates may enter, exit, or become inactive between those proposals
and the collection date.

This temporal mismatch can produce values that do not represent a historical
proposal-specific participation fraction. Observations exceeding one are
therefore treated as unavailable rather than forced into the analysis.

The measure also gives equal address-level weight to delegates regardless of
their voting power and does not capture deliberation quality, independence,
voting rationale, or delegator representation.

Supply-normalized voting turnout provides a complementary measure based on
voting power cast relative to governance-token supply. However, this is not the
same as turnout relative to contemporaneously delegated or eligible voting
power. Participation should therefore remain a secondary, multidimensional
construct rather than a single definitive measure.

\subsection{Effective Quorum Coalition}
\label{sec:limitations-quorum}

The Effective Quorum Coalition combines the current observed delegate
distribution with a quorum threshold constructed from recent governance
records. The quorum value is based primarily on the median of up to the five
most recent concluded proposals with usable information, with a governor-level
value used where necessary.

This provides a more stable cross-sectional representation than relying on a
single proposal, but it is not a historical proposal-specific coalition
measure. Both voting-power distribution and quorum parameters may have differed
for earlier proposals.

A valid coalition size is observed only where delegate voting power and quorum
are expressed in compatible units and the currently delegated voting power
reaches the relevant threshold. Eleven DAOs have current observed delegation
below the reconstructed quorum, two contain implausible quorum units, two lack
usable delegation information, and one lacks a usable quorum value.

These cases should not be interpreted as exact large coalition sizes. In
particular, ``unreachable'' means only that the observed current delegation is
insufficient to reach the recent quorum threshold; additional delegation or a
different governance state could change that status.

Finally, $Q_i$ establishes arithmetic sufficiency for quorum, not proposal
passage. A value of $Q_i=1$ therefore indicates SDQS, not unilateral governance
control.

\subsection{Token-Market Scale Measurement and Endogeneity}
\label{sec:limitations-endogeneity}

Live FDV, CryptoCompare market capitalization, and trading-life-average FDV are
token-market valuation measures rather than comprehensive measures of
organizational size. They do not directly capture treasury assets, protocol
revenue, transaction activity, developer activity, user adoption, or other
dimensions of organizational scale.

The primary live-FDV sample is also relatively small, which limits statistical
power. For the primary HHI rank analysis with $N=28$, the approximate minimum
detectable absolute correlation at 80\% power is about 0.51. Failure to reject
zero should therefore be interpreted as an absence of a statistically
detectable robust relationship in the present sample rather than evidence that
moderate effects are impossible.

The observed positive associations between token-market scale and delegate or
holder counts are also cross-sectional and potentially endogenous. Larger,
older, or more successful protocols may attract more governance participants,
while broader governance participation may itself contribute to organizational
development. The present design does not identify either direction.

\subsection{Crypto-Market Benchmark}
\label{sec:limitations-benchmark}

Bitcoin provides a common and transparent benchmark for systematic
crypto-market exposure, but it is not a complete crypto-asset pricing model.
BTC-relative returns may retain exposure to market-wide, sector-specific,
liquidity, momentum, or token-specific factors.

Accordingly, the quantity

\begin{equation}
    r_{it}-r_{\mathrm{BTC},t}
\end{equation}

should be interpreted as a BTC-relative return rather than a fully
risk-adjusted abnormal return. The strong estimated Bitcoin exposure in the
monthly panel demonstrates substantial contemporaneous co-movement but does not
imply that Bitcoin is the only systematic factor affecting DAO-token returns.

\subsection{Proposal-Event Identification}
\label{sec:limitations-event-overlap}

The proposal-event design uses proposal completion as a standardized governance
timestamp. This choice permits comparable event alignment but does not imply
that the proposal end date is the moment at which new information reaches the
market. Proposal content, delegate positions, voting trajectories, and likely
outcomes may already be public before voting closes.

Repeated and overlapping events also complicate inference. The revised analysis
addresses this issue using distinct DAO--event dates, non-overlapping within-DAO
events, DAO clustering, DAO--calendar-month clustering, DAO-level aggregation,
and placebo windows. The weakening of statistical significance under these
specifications is itself informative: inference about proposal-related returns
is sensitive to the dependence structure of governance events.

A further limitation is proposal heterogeneity. The current analysis pools
routine parameter changes, treasury decisions, upgrades, governance-process
changes, and potentially security-relevant proposals. If these proposal types
have different informational content, their aggregation may attenuate or mask
type-specific effects. A reliable proposal classification is therefore an
important extension for future work.

\subsection{Incident Evidence}
\label{sec:limitations-incidents}

The current governance-under-stress analysis uses a small set of illustrative
incidents rather than a systematically enumerated incident population.
Accordingly, the cases can demonstrate mechanisms and boundary conditions but
cannot establish the frequency of particular failure modes or the comparative
effectiveness of emergency-governance designs.

A systematically collected incident corpus with explicit inclusion criteria and
common coding dimensions would be required before making population-level
claims about governance capture, response latency, emergency authority, or
post-incident accountability.

\section{Conclusion}
\label{sec:conclusion}

This study develops an integrated framework for understanding DAO governance by
connecting formal governance design, the realized distribution of decision
authority, observable economic outcomes, and governance under stress. The
resulting perspective moves beyond treating decentralization as a property of
token voting alone. Agenda access, voting and delegation, decision thresholds, execution constraints, emergency authority, and external dependencies jointly determine where governance power resides and how it can be exercised.

The empirical evidence demonstrates why this distinction matters. Within the
observed Tally-indexed DAO sample, broad formal participation opportunities
frequently coexist with concentrated effective authority. The median largest
delegate controls approximately 42\% of delegated voting power, the ten largest
delegates collectively control approximately 98\%, and the median governance
Nakamoto coefficient is two. Among DAOs for which current delegated voting
power reaches a valid quorum, the median Effective Quorum Coalition is one,
with approximately 63\% exhibiting single-delegate quorum sufficiency. These
findings do not imply collusion, governance capture, or unilateral proposal
passage. They instead show that nominal openness and effective distribution of
authority are empirically distinct properties.

The economic analysis reinforces this multidimensional interpretation.
Token-market scale is more consistently associated with the breadth of the
delegate and token-holder bases than with an independently less concentrated
distribution of voting power. Static governance characteristics do not robustly
explain average token returns, while contemporaneous crypto-market movements
account for substantially more monthly return variation. Governance proposal
events produce negative BTC-relative point estimates of approximately
1.5--2\% over the $[-5,+5]$ window after accounting for DAO-specific non-event
performance, but this evidence weakens when repeated events, common calendar
dependence, and overlapping windows are addressed. Governance is therefore
institutionally consequential without necessarily providing a simple or robust
short-run token-return signal.

The broader implication is that DAO decentralization should be evaluated as a
\emph{configuration of authority} rather than as a binary label or a single
score. Formal governance determines where authority can reside; effective
decentralization reveals where it actually resides; stress conditions reveal
how that authority can be exercised when ordinary procedures become
insufficient. Accordingly, the relevant questions are not simply whether a DAO
is decentralized, but \emph{who can place issues on the agenda, who controls
effective voting power, who participates, how broadly authority must be combined
to satisfy decision thresholds, who ultimately executes approved actions, and
who can intervene when the system is under stress}.

By linking these questions through a common
design--measurement--outcome framework, this study provides a basis for
evaluating DAO governance without conflating transparency with
decentralization, structural exposure with realized capture, or governance
quality with token-price performance. The central conclusion is therefore
simple: \textbf{the meaningful unit of DAO decentralization is not the
availability of governance rights, but the distribution, exercise, and
constraint of effective authority across the governance lifecycle.}

\section*{Declaration of competing interest}

The authors declare that they have no known competing financial interests or
personal relationships that could have appeared to influence the work
reported in this paper.

\section*{Funding}

This research did not receive any specific grant from funding agencies in the
public, commercial, or not-for-profit sectors.

\section*{Data and code availability}

The replication package submitted as supplementary material contains cached
source responses, inclusion and exclusion records, token-matching logs,
processed datasets, analysis scripts, generated tables, and figure inputs.
Data obtained from public third-party services remain subject to the
applicable providers' terms and availability.

\section*{Declaration of generative AI and AI-assisted technologies in the
writing process}

During the preparation of this work, the authors used OpenAI's ChatGPT for
language editing, manuscript organization, and \LaTeX{} drafting. The authors
reviewed and edited all resulting material and take full responsibility for the
content of the publication.

\section*{CRediT authorship contribution statement}

\textbf{Yash Madhwal:} Conceptualization, Methodology, Software, Validation,
Formal analysis, Investigation, Data curation, Visualization, Writing --
original draft, Writing -- review \& editing, Project administration.
\textbf{Vladimir Lebedev:} Conceptualization, Methodology, Formal analysis,
Investigation, Writing -- original draft, Writing -- review \& editing.
\textbf{Alexey Karanyuk:} Methodology, Validation, Investigation, Writing --
review \& editing. \textbf{Yury Yanovich:} Conceptualization, Methodology,
Supervision, Project administration, Writing -- review \& editing.

\bibliographystyle{elsarticle-num}
\bibliography{mybibliography}

\clearpage
\onecolumn
\appendix
\section{Structured SoK Evidence Matrix}
\label{app:sok-evidence-matrix}

Tables~\ref{tab:sok-matrix-a}--\ref{tab:sok-matrix-c} report the complete core
academic corpus used in the concept-centric SoK. The matrix records the
evidentiary basis of the synthesis: each coded study is mapped to a governance
dimension, method, scope, evidence type, and principal finding. The
``evidence'' column identifies the material analyzed by each study, while the
final column records the finding coded for the present governance framework.

\begin{table}[htbp]
\centering
\caption{Core SoK evidence matrix, Part I: foundational and conceptual studies.}
\label{tab:sok-matrix-a}
\footnotesize
\setlength{\tabcolsep}{2pt}
\renewcommand{\arraystretch}{1.13}
\begin{tabular}{@{}p{0.115\textwidth}p{0.04\textwidth}p{0.12\textwidth}p{0.125\textwidth}p{0.105\textwidth}p{0.15\textwidth}p{0.235\textwidth}@{}}
\hline
\textbf{Study} & \textbf{Year} & \textbf{Governance dimension} &
\textbf{Method} & \textbf{DAO(s) / scope} & \textbf{Evidence} &
\textbf{Main finding coded in this SoK} \\
\hline
De Filippi \& Loveluck \cite{defilippi2016politics} & 2016 & Off-chain power &
Interpretive case study & Bitcoin & Block-size dispute and public records &
Technical decentralization does not remove elite politics or informal authority. \\
Hsieh et al. \cite{hsieh2018bitcoin} & 2018 & Organizational design &
Conceptual case analysis & Bitcoin & Protocol and organizational roles &
Blockchain systems enable a distinct organizational form, but coordination roles remain. \\
Beck et al. \cite{beck2018blockchain} & 2018 & Governance layers &
Conceptual framework & Blockchain ecosystems & Literature and theory &
Governance operates through interacting decision rights, accountability, and incentives. \\
Azouvi et al. \cite{azouvi2019governance} & 2019 & Code governance &
Quantitative mapping & Bitcoin and Ethereum & Repositories and contributor data &
Technical governance contributions are concentrated despite open participation. \\
El Faqir et al. \cite{elfaqir2020overview} & 2020 & DAO platforms &
Comparative overview & Aragon, DAOstack, DAOhaus, Colony & Platform documents and tooling &
DAO platforms implement heterogeneous governance mechanisms and organizational services. \\
De Filippi et al. \cite{defilippi2020confidence} & 2020 & Trust and accountability &
Conceptual synthesis & Blockchain systems & Legal and governance theory &
Code can create procedural confidence but does not eliminate trust or governance challenges. \\
Hassan \& De Filippi \cite{hassan2021dao} & 2021 & DAO definition &
Concept analysis & DAO ecosystem & Literature and deployed systems &
DAOs combine autonomous rules with socially organized collective action. \\
Honkanen et al. \cite{honkanen2021buildingblocks} & 2021 & Roles and incentives &
Content analysis & 241 blockchain projects & White-paper coding &
Recurring organizational building blocks distribute governance roles and incentives. \\
Bod\'{o} et al. \cite{bodo2021decentralisation} & 2021 & Decentralization dimensions &
Multidisciplinary synthesis & Distributed systems & Technical, social, political, and economic literature &
Network topology alone cannot proxy the distribution of authority. \\
Rozas et al. \cite{rozas2021ostrom} & 2021 & Commons governance &
Theory-led case analysis & Blockchain commons and peer production & Cases mapped to Ostrom's principles &
Blockchain affordances can support, but do not guarantee, collective governance. \\
Reijers et al. \cite{reijers2021code} & 2021 & On-/off-chain governance &
Legal-philosophical case analysis & The DAO and Ethereum & Protocol and social-response records &
On-chain code remains embedded in off-chain institutions and intervention. \\
van Pelt et al. \cite{vanpelt2021framework} & 2021 & Governance taxonomy &
Design-science framework & Blockchain systems & Literature synthesis and validation &
Three governance layers and six dimensions support structured comparison. \\
\hline
\end{tabular}
\end{table}

\clearpage
\begin{table}[htbp]
\centering
\caption{Core SoK evidence matrix, Part II: governance design and review studies.}
\label{tab:sok-matrix-b}
\footnotesize
\setlength{\tabcolsep}{2pt}
\renewcommand{\arraystretch}{1.13}
\begin{tabular}{@{}p{0.115\textwidth}p{0.04\textwidth}p{0.12\textwidth}p{0.125\textwidth}p{0.105\textwidth}p{0.15\textwidth}p{0.235\textwidth}@{}}
\hline
\textbf{Study} & \textbf{Year} & \textbf{Governance dimension} &
\textbf{Method} & \textbf{DAO(s) / scope} & \textbf{Evidence} &
\textbf{Main finding coded in this SoK} \\
\hline
Liu et al. \cite{liu2022principles} & 2022 & Design principles &
Framework and case studies & Five blockchain platforms & Literature and case evidence &
Decentralization, rights, incentives, and accountability must be assessed jointly. \\
Santana \& Albareda \cite{santana2022dao} & 2022 & Organizational model &
Integrative review and model & DAOs generally & Literature synthesis &
Decentralization, automation, and autonomous community jointly characterize DAOs. \\
Arroyo et al. \cite{arroyo2022daoAnalyzer} & 2022 & Activity and participation &
Visual-analytics tool study & Platform-indexed DAOs & Proposal and vote time series &
Longitudinal activity and participation require joint inspection. \\
Appel \& Grennan \cite{appel2023control} & 2023 & Control &
Empirical organizational analysis & Multiple DAOs & Proposal and voting records &
Formal openness can coexist with control by a small set of influential entities. \\
Sch\"{a}dler et al. \cite{schadler2023decisionmaking} & 2023 & Decision rights &
Exploratory multi-case study & Five blockchain systems & Documents, interviews, and case evidence &
Decision-making varies by community versus institution and on-chain versus off-chain locus. \\
Rikken et al. \cite{rikken2023insouts} & 2023 & Definition and viability &
Grounded-theory literature coding & On- and off-chain DAOs & Coded definitions and studies &
Objectives, infrastructure, and users condition DAO viability. \\
Rikken et al. \cite{rikken2023impacts} & 2023 & Accountability and incentives &
Empirical comparative analysis & Observed DAOs & Case and documentary evidence &
Governance elements shape organizational impacts and long-term viability. \\
Zhu et al. \cite{zhu2024governance} & 2024 & Governance technology &
Structured survey & Blockchain systems & Technical literature &
On-chain and off-chain mechanisms address different governance failure modes. \\
Gilson \& Bouraga \cite{gilson2024democratic} & 2024 & Voting design &
Qualitative analysis & DAO voting systems & Literature and mechanism comparison &
Liquid and quadratic voting involve implementation-specific democratic trade-offs. \\
van Vulpen et al. \cite{vanvulpen2024oss} & 2024 & Open-source governance &
Design framework and case synthesis & OSS-producing DAOs & OSS and DAO governance evidence &
Governance design must fit software-production roles and contribution processes. \\
Weidener et al. \cite{weidener2025delegated} & 2025 & Delegation &
Systematic scoping review & Delegated-voting DAOs & Academic and implementation literature &
Delegation can reduce participation burdens while increasing concentration and accountability risks. \\
\hline
\end{tabular}
\end{table}

\clearpage
\begin{table}[htbp]
\centering
\caption{Core SoK evidence matrix, Part III: empirical, security, and economic studies.}
\label{tab:sok-matrix-c}
\footnotesize
\setlength{\tabcolsep}{2pt}
\renewcommand{\arraystretch}{1.13}
\begin{tabular}{@{}p{0.115\textwidth}p{0.04\textwidth}p{0.12\textwidth}p{0.125\textwidth}p{0.105\textwidth}p{0.15\textwidth}p{0.235\textwidth}@{}}
\hline
\textbf{Study} & \textbf{Year} & \textbf{Governance dimension} &
\textbf{Method} & \textbf{DAO(s) / scope} & \textbf{Evidence} &
\textbf{Main finding coded in this SoK} \\
\hline
Gudgeon et al. \cite{gudgeon2020crisis} & 2020 & Capture and resilience &
Constructive attack and stress analysis & MakerDAO and DeFi lending & Protocol parameters and market scenarios &
Governance manipulation and liquidity shocks can compound protocol risk. \\
Eskandari et al. \cite{eskandari2021sokOracles} & 2021 & External dependencies &
SoK and taxonomy & Oracle systems & Designs, attacks, and incidents &
Oracle design creates manipulation, liveness, and governance exposure. \\
Goldberg \& Sch\"{a}r \cite{goldberg2023metaverse} & 2023 & Participation and voter behavior &
On-chain empirical analysis & Decentraland & Governance votes and addresses &
Participation and voting power are unevenly distributed. \\
Fritsch et al. \cite{fritsch2024votingpower} & 2024 & Voting power and delegation &
On-chain empirical analysis & Compound, Uniswap, ENS & Holders, delegates, proposals, and votes &
Voting power is concentrated among a limited set of addresses. \\
Feichtinger et al. \cite{feichtinger2024shortcomings} & 2024 & Concentration, cost, and activity &
On-chain empirical study & 21 DAOs & Live governance records &
Concentration, hidden governance costs, and low-value activity recur across systems. \\
Pe\~{n}a-Calvin et al. \cite{penacalvin2024concentration} & 2024 & Participation and concentration &
Ecosystem-scale empirical analysis & Multi-platform DAO ecosystem & Proposals, voters, and voting power &
Nominal openness coexists with low participation and concentrated power. \\
Sun et al. \cite{sun2024makerdao} & 2024 & Centralization and outcomes &
Empirical time-series analysis & MakerDAO polls & Polls plus MKR, DAI, network, and sentiment data &
Governance centralization is associated with performance trade-offs. \\
Feichtinger et al. \cite{feichtinger2024sokdaos} & 2024 & Attack surface &
SoK and attack taxonomy & Multiple DAOs and incidents & Documented and constructive attacks &
DAO attack vectors span technical, economic, and governance layers. \\
Bellavitis \& Momtaz \cite{bellavitis2025voting} & 2025 & Voting and value creation &
Empirical governance-market analysis & Investment DAOs & Proposal, vote, and performance data &
Voting information and governance design are associated with value creation. \\
Lo Monaco et al. \cite{lomonaco2025distributed} & 2025 & Distributed decisions and value &
Regression discontinuity design & Contested DAO proposals & Close-vote outcomes and token value &
Distributed governance can increase tokenholder value when decisions are contested. \\
Han et al. \cite{han2026dao} & 2026 & Whale conflict and incentives &
Analytical model with empirical implications & Stylized DAOs and platforms & Token trading, voting, and staking mechanisms &
Liquidity and token lock-in shape whale--minority conflicts and platform growth. \\
\hline
\end{tabular}
\end{table}

\clearpage

\section{Search Protocol and Sample Distinctions}
\label{app:search-protocol}

The paper uses two distinct evidentiary objects. The SoK corpus is a literature corpus: it contains 34 peer-reviewed journal or conference studies that were coded for governance mechanisms, boundary conditions, and trade-offs. The empirical sample is a DAO corpus: it contains 54 organizations identified through Tally and screened using the eligibility rules reported in Section~\ref{sec:empirical-data}. These two counts should not be conflated: the former indexes scholarly evidence, whereas the latter indexes deployed governance systems.

The SoK retrieval was concept-centric. It began from foundational blockchain-governance and DAO-governance frameworks and was expanded through backward and forward citation chaining. Targeted searches combined ``DAO'' or ``decentralized autonomous organization'' with ``governance'', ``voting'', ``delegation'', ``participation'', ``quorum'', ``voting power'', ``capture'', ``attack'', ``emergency'', ``oracle'', ``valuation'', and ``returns''. Records were considered through the data-collection cut-off used in the empirical analysis.

A source entered the core coding corpus when it contributed one of the following: a governance model or design taxonomy; an empirical result on voting, delegation, participation, or concentration; a security, attack, or failure analysis; or evidence linking governance arrangements to organizational or market outcomes. Sources were excluded when they only mentioned DAOs without contributing a coded governance mechanism, duplicated a stronger version of the same result, or lacked sufficient detail to support a coded claim. Citation chaining was stopped when additional sources reproduced an already represented mechanism without changing the taxonomy or its boundary conditions.

The DAO sample was constructed from Tally-indexed organizations rather than from a manually selected set of prominent DAOs. Tally was the source of governance artifacts, and CryptoCompare was the source of token-price data. An organization was retained only if it satisfied the mainnet governance-token, non-duplication, proposal-history, and token-matching rules described in Section~\ref{sec:empirical-data}. This procedure explains why the empirical DAO count is 54, whereas the coded SoK literature count is 34.

\end{document}